\documentclass[aps,pra,reprint,longbibliography]{revtex4-2}

\usepackage{amsfonts,amscd,mathrsfs,amsmath,amsthm,amssymb}
\usepackage{mathtools}
\usepackage{booktabs}
\usepackage{physics}
\usepackage{xparse}
\usepackage{graphicx}
\usepackage{float}
\usepackage{pifont}
\usepackage[dvipsnames]{xcolor}
\usepackage{colortbl}
\usepackage{multirow}
\usepackage{enumerate}   
\usepackage{comment}
\usepackage{bm}
\usepackage{bbm}
\usepackage{caption}
\usepackage{subcaption}
\usepackage{adjustbox}
\PassOptionsToPackage{hyphens}{url}\usepackage{hyperref}
\hypersetup{colorlinks=true,linkcolor=blue!80!black,  citecolor=blue!80!black,}
\usepackage[capitalise]{cleveref}

\newcommand\numberthis{\addtocounter{equation}{1}\tag{\theequation}}

\newcommand{\sm}{\text{-}}
\newcommand{\CNOT}{\mathrm{CNOT}}

\makeatletter
\newcommand*{\rom}[1]{\expandafter\@slowromancap\romannumeral #1@}
\makeatother



    \DeclareMathAlphabet{\mathsfit}{T1}{\sfdefault}{\mddefault}{\sldefault}
    \SetMathAlphabet{\mathsfit}{bold}{T1}{\sfdefault}{\bfdefault}{\sldefault}

    \newcommand{\U}{\mathrm{U}}
    \newcommand{\SO}{\mathrm{SO}}
        \newcommand{\GL}{\mathrm{GL}}
    \newcommand{\Q}{\mathsf{Q}}

    \newcommand{\K}{\mathsf{K}}
    
    \newcommand{\F}{\mathsf{F}}
    
    \newcommand{\BD}{\mathsf{BD}}

      \newcommand{\tet}{\mathsf{2T}}
    \newcommand{\oct}{\mathsf{2O}}
    \newcommand{\ico}{\mathsf{2I}}

    \newcommand{\G}{\mathsf{G}}

\newcommand{\ZZ}{\mathbb{Z}}

\newcommand{\X}{\mathsf{X}}
\newcommand{\Y}{\mathsf{Y}}
\newcommand{\Z}{\mathsf{Z}}
\renewcommand{\F}{\mathsf{F}}
\renewcommand{\H}{\mathsf{H}}
\renewcommand{\S}{\mathsf{S}}
\newcommand{\Ph}{\mathsf{Ph}}

\newcommand{\M}{\mathsf{M}}
\newcommand{\N}{\mathsf{N}}

\renewcommand{\P}{\mathsf{P}}
\newcommand{\C}{\mathsf{C}}
\newcommand{\Ex}{\mathsf{Ex}}

\newcommand{\SU}{\mathrm{SU}}
\newcommand{\PU}{\mathrm{PU}}
\newcommand{\Sp}{\mathrm{Sp}}
\newcommand{\SL}{\mathrm{SL}}
\newcommand{\Alt}{\mathrm{A}}
\newcommand{\Sym}{\mathrm{S}}

\renewcommand{\CNOT}{\mathsf{CNOT}}
\newcommand{\BELL}{\mathsf{BELL}}
\newcommand{\Swap}{\mathsf{SWAP}}

\newcommand{\CZ}{\mathsf{CZ}}
\newcommand{\DCNOT}{\mathsf{DCNOT}}

\newcommand{\I}{\mathsf{I}}

\newcommand\otimesswap{%
  \mathrel{\ooalign{\hss\hbox{\scalebox{0.9}{$\otimes$}}\hss\cr%
  \kern0.2ex\raise1.4ex\hbox{\scalebox{0.6}{$\leftrightarrows$}}}}}

\begin{document}

\title{Classification of the Subgroups of the Two-Qubit Clifford Group}
\thanks{These authors contributed equally to this work.}
\author{Eric Kubischta}
\email{erickub@umd.edu} 
\author{Ian Teixeira}
\email{igt@umd.edu}

\affiliation{Joint Center for Quantum Information and Computer Science,
NIST/University of Maryland, College Park, Maryland 20742 USA}

\begin{abstract}
We perform a complete classification of all $ 56 $ subgroups of the two-qubit Clifford group containing the two-qubit Pauli group. We provide generators for these groups using gates familiar to the quantum information community and we reference these groups against the group libraries provided in GAP. We also list several families of groups in higher levels of the two-qubit Clifford hierarchy.
\end{abstract}

\maketitle

\section{Introduction}

The transversal gates of a quantum error-correcting code are naturally fault tolerant and form a group $\G$, which must be finite by the Eastin-Knill theorem \cite{eastinKnill}. For an $ [[n,k,d]] $ stabilizer code, $ \G $ must be contained in some finite level of the $ k $-qubit Clifford hierarchy \cite{disjointnessStabilizerCodes}, and moreover must contain the $ k $-qubit Pauli group \cite{stabilizerThesis}.

For an $ [[n,1,d]] $ stabilizer code, the possible groups $ \G $ are well known due to the simplicity of the classification of the finite subgroups of $\SU(2)$. For $r\geq 3$, the only group in the $r$-th level of the single-qubit Clifford hierarchy that also contains the Pauli group is the generalized quaternion group $\Q^{(r)}$. The only groups in the 2nd level of the single-qubit Clifford hierarchy that contain the Pauli group are: (1) the single qubit Clifford group $\C_1$ (also known as the binary octahedral group $\oct$), and (2) the group $\C_1'$ (the commutator subgroup of $\C_1$ also known as the binary tetrahedral group $\tet$), and (3) the generalized quaternion group $\Q^{(2)}$ consisting of the Paulis with the phase gate $\S$. 

For an $ [[n,2,d]] $ stabilizer code, the story is much less clear since the finite subgroups of $ \SU(4) $ are not well understood. A partial classification of the finite subgroups of $\SU(4)$ was given in 1917 \cite{blichfeldt1917finite} and the results were recapitulated in modern notation in 1999 \cite{subgroupsSU4}. But these papers contain several errors, are seemingly incomplete (with regards to the monomial and non-monomial sections), and are not written using gates familiar to the quantum information community. Moreover, it is not clear from either of these two papers how the subgroups of $ \SU(4)$ are related to the two-qubit Clifford hierarchy.

Due to the incomplete state of the current literature \cite{blichfeldt1917finite,subgroupsSU4} we won't be able to classify all groups in the two-qubit Clifford hierarchy. However, we are able to classify for the first time all subgroups of the two-qubit Clifford group containing the Pauli group. We organize the groups based on their entanglement structure and we write down generators of these groups using familiar gates. In the appendix we will list many infinite series of groups that appear in higher levels of the Clifford hierarchy.

\section{Background}

\subsection{Single-Qubit Gates}
Quantum gates are usually presented as elements of the unitary group. However, $\U(2^k) = e^{i \theta} \SU(2^k)$ and so it is sufficient to consider gates in $\SU(2^k)$ instead since global phase is irrelevant to quantum. To aid in this distinction, we will write gates in $\SU(2^k)$ using sans serif font. For example, the phase gate in $\SU(2)$ is denoted by $\S$ instead of the $\U(2)$ version $S$. 

The Pauli gates are $I = \smqty(1 & 0 \\ 0 & 1)$, $X = \smqty(0 & 1 \\ 1 & 0)$, $Y = \smqty( 0 & \sm i \\ i & 0)$, and $Z = \smqty(1 & 0 \\ 0 & \sm 1)$. The determinant-1 versions of these gates are $\I := I$, $\X := \sm i X$, $\Y := \sm i Y$, and $\Z := \sm i Z$ and they generate the determinant-1 single qubit Pauli group which we denote as $\P_1$.

The single-qubit Clifford group normalizes the Pauli group. Common Clifford gates include the phase gate $S = \smqty(1 & 0 \\ 0 & i)$, the Hadamard gate $H = \tfrac{1}{\sqrt{2}}\smqty(1 & 1 \\ 1 & \sm 1)$, and the Facet gate $F = \tfrac{1}{\sqrt{2}}\smqty(1 & \sm i \\ 1 & i)$ \footnote{See the Supplemental Material of \cite{us1} for a history of the Facet gate.}. The determinant-1 versions of these gates are $\H = \sm i H$, $\S = \zeta_8^* S$, and $\F = \zeta_8^* F$ where $\zeta_m := e^{2\pi i/m}$ is an $m$-th root of unity and $*$ denotes complex conjugation (here and throughout the paper). We denote the determinant-1 (special) single qubit Clifford group by $\C_1$ and it contains the Pauli group $\P_1$ as well as $\S$, $\H$, and $\F$.  

The largest subgroup of the single qubit Clifford group $\C_1$ is the commutator subgroup $\C_1' := [ \C_1, \C_1]$, which contains the Pauli group $\P_1$ as well as the facet gate $\F$ (but not $\H$ or $\S$). Recall that, given a group $ G $, the commutator subgroup $ G' $ is the group generated by $ g_1g_2g_1^{-1}g_2^{-1}$ for all $ g_1,g_2 \in G $. Intuitively, $G'$ is the smallest normal subgroup of $G$ such that $G/G'$ is abelian.

Finally, we can consider the generalized phase gate
$
\Ph(\tfrac{2\pi}{m})=\smqty( e^{\sm i \pi/m} & 0 \\ 0 & e^{i \pi/m} )
$
which generalizes $ \Z=\Ph(\tfrac{2\pi}{2}) $ and $ \S=\Ph(\tfrac{2\pi}{4}) $. This diagonal gate $ \Ph(\tfrac{2\pi}{m}) $ together with $ \X $ generates the degree $ m $ binary dihedral group
$
\BD_{m}=\expval{\Ph(\tfrac{2\pi}{m}),\X}.
$
When $m$ is a power of $2$, i.e., $ m=2^r $, the group $ \BD_{2^r}$ is also called the generalized quaternion group $\Q^{(r)}$ and is contained in the $ r $-th level of the single qubit Clifford hierarchy \cite{us2}. For example the groups $ \Q^{(2)}=\expval{\X,\S} $ and the Pauli group $\Q^{(1)}=\expval{\X,\Z}=\P_1 $ are contained in $ \C_1 $.  

\subsection{Two-Qubit Gates}

Local gates in $\SU(4)$ are of the form $\mathsf{U} \otimes \mathsf{V}$ where $\mathsf{U},\mathsf{V} \in \SU(2)$. For example, the two-qubit Pauli group is defined to be $\P_2 := \P_1 \otimes \P_1$, that is, it consists of local Pauli gates. 

The two-qubit computational basis is defined as $\ket{00}$, $\ket{01}$, $\ket{10}$, and $\ket{11}$. A Bell basis consists of the maximally entangled two-qubit Bell states $\tfrac{1}{\sqrt{2}}\qty(\ket{00} + \ket{11})$, $\tfrac{1}{\sqrt{2}}\qty( \ket{00} - \ket{11})$, $\tfrac{1}{\sqrt{2}}\qty(\ket{01} + \ket{10})$, and $\tfrac{1}{\sqrt{2}}\qty(\ket{01} - \ket{10})$, with any relative phases between these four states. It was shown in \cite{BellGate1997,BellGate2004} that a particular choice of Bell basis, the columns of the \textit{Bell gate}
\[
\BELL =  \tfrac{\zeta_8^3}{\sqrt{2}} \smqty( 1 & i & 0 & 0 \\ 0 & 0 & i & 1 \\ 0 & 0 & i & -1 \\ 1 & -i & 0 & 0), \numberthis
\]
yields a matrix transformation that conjugates any real matrix in $ \SU(4) $ to a tensor product of two single qubit gates. In other words, $ \BELL $ conjugates the subgroup $ \SO(4) $ to the subgroup $ \SU(2) \otimes \SU(2) $. Note that $ \BELL $ is a Clifford gate and appears in some familiar contexts. For example, $ \BELL^{\otimes 7} $ implements logical $\BELL^*$ for two blocks of the $[[7,1,3]]$ Steane code.

A monomial matrix is the product of a permutation matrix and a diagonal matrix. Many common quantum gates are monomial, for example all Pauli gates. Another example is the two-qubit swap gate, whose determinant-$ 1 $ version is 
\[
\Swap =  \zeta_8^* \smqty( 1 & 0 & 0 & 0 \\ 0 & 0 & 1 & 0 \\ 0 & 1 & 0 & 0 \\ 0 & 0 & 0 & 1). \numberthis
\]
The swap gate is a key example of a non-entangling Clifford gate. Another nonentangling gate is the scalar matrix 
\[
i\I\I=\smqty( i & 0 & 0 & 0 \\ 0 & i & 0 & 0 \\ 0 & 0 & i & 0 \\ 0 & 0 & 0 & i) \numberthis
\]
which just applies a global phase.

Now we list some important entangling Clifford gates. First we have the (special) controlled-not, with the second qubit as its target,
\[
\CNOT_{12}= \zeta_8^* \smqty(1 & 0 & 0 & 0\\ 0 & 1 & 0 & 0\\0 & 0 & 0 & 1\\0 & 0 & 1 & 0\\). \numberthis
\]
We also have the (special) controlled-not gate that has the first qubit as its target,
\[
\CNOT_{21}= \zeta_8^* \smqty(1 & 0 & 0 & 0\\ 0 & 0 & 0 & 1\\0 & 0 & 1 & 0\\0 & 1 & 0 & 0\\). \numberthis
\]
Sometimes we will drop the subscripts for brevity and in those cases we will always mean the standard $\CNOT_{12}$ version, i.e., $\CNOT := \CNOT_{12}$.

The double controlled-not is just a product of the two different controlled-not gates. Note that the double controlled-not gate is naturally determinant $ 1 $ so we are not doing anything special by taking  the determinant $ 1 $ version:
\[
\DCNOT= \mathrm{CNOT}_{12} \cdot \mathrm{CNOT}_{21}= \smqty(1 & 0 & 0 & 0\\ 0 & 0 & 0 & 1\\0 & 1 & 0 & 0\\0 & 0 & 1 & 0\\). \numberthis
\]

Lastly we will need the determinant-1 controlled-$ Z $ gate:
\[
\CZ:=\zeta_8^* \smqty(1 & 0 & 0 & 0\\ 0 & 1 & 0 & 0\\0 & 0 & 1 & 0\\0 & 0 & 0 & \sm 1\\). \numberthis 
\]

\subsection{The Clifford Hierarchy}

The two-qubit (determinant-1) Clifford hierarchy \cite{cliffordhierarchy} is defined recursively as
\[
    \C^{(r)}_2 := \{ \mathsf{U} \in \SU(4) : \mathsf{U} \P_2 \mathsf{U}^
    \dagger \in \C^{(r-1)}_2 \}, \numberthis
\]
starting with $\C_2^{(1)} :=  \P_2$, the two-qubit Pauli group. The second level $\C_2^{(2)}$ is the two-qubit Clifford group $\C_2$, but $ \C^{(r)}_2 $ is not a group for $ r \geq 3 $ . As in \cite{us1}, we will call a gate \textit{exotic} if it is not in any level of the Clifford hierarchy, and we will call a group exotic if its contains exotic gates. 

\subsection{Irreducible Groups}
A group $G$ acts reducibly on the Hilbert space $\mathcal{H}$ if we can write $\mathcal{H} = \mathcal{H}_1 \oplus \cdots \oplus \mathcal{H}_n$ such that every $g \in G$ fixes each subspace $\mathcal{H}_i$. Specifically, $g \cdot \mathcal{H}_i = \mathcal{H}_i$ for all $g \in G$ and all $\mathcal{H}_i$. So a reducible group splits the total Hilbert space up into more than one sector. An \textit{irreducible} group is one that is not reducible, meaning that the Hilbert space does not split into more than one sector.

Since the transversal gate group $ \G $ of an $ [[n,k,d]] $ stabilizer code always contains the $ k $-qubit Pauli group $ \P_k $, and $ \P_k $ is irreducible as a subgroup of $ \SU(2^k) $, then $ \G $ must also be irreducible. Thus in our analysis of Clifford subgroups it is sufficient to restrict our attention to subgroups that are irreducible.

\subsection{Primitive Groups}

For an irreducible group $G$ we cannot split the Hilbert space as $\mathcal{H} = \mathcal{H}_1 \oplus \mathcal{H}_2$ such that $g \cdot \mathcal{H}_i = \mathcal{H}_i$. But suppose it was the case that $g \cdot \mathcal{H}_1 = \mathcal{H}_2$ and $g \cdot \mathcal{H}_2 = \mathcal{H}_1$. This looks almost like a reducible action but with a permutation. In this sense, an irreducible group can still reduce the Hilbert space into pseudo-sectors. A group is called \text{primitive} \cite{blichfeldt1917finite} if it does not split the Hilbert space into more than one pseudo-sector. A group is called imprimitive if it is irreducible but not primitive. Primitive groups can be considered even more basic than irreducible groups.

\subsection{Outline}

There are $ 56 $ subgroups of the Clifford group $ \C_2 $ containing the two-qubit Pauli group. This can be determined by using GAP \cite{GAP} and calling the \texttt{IntermediateSubgroups} command which produces a list of 1453 proper subgroups of $ \C_2 $ containing the Pauli group. From there it can be determined that there are only $ 56 $ subgroups on this list up to isomorphism (including the Clifford group and the Pauli group). We have then sorted through and classified these subgroups, finding familiar generators and elucidating interesting properties of each group.

There are $ 17 $ primitive finite subgroups of $ \SU(4) $ containing $\P_2$, and all of these groups are contained in the two-qubit Clifford group $\C_2$. Note that these are the only primitive subgroups of $ \C_2 $ that can arise as the transversal gate group of an $ [[n,2,d]] $ stabilizer code, since the transversal gate group must contain $ \P_2 $. Of these $ 17 $ groups, $ 4 $ consist of only local gates, $ 4 $ are non-entangling, and $ 9 $ contain entangling gates. We describe these primitive groups in our first section. 

A monomial matrix is the product of a permutation matrix and a diagonal matrix and a monomial group is a group of monomial matrices. In our second section we describe the monomial subgroups of $ \C_2 $, which are imprimitive since up to relative phase they just permute the four coordinate axes in $ \mathbb{C}^2 \otimes \mathbb{C}^2 \cong \mathbb{C}^4$. Then in our third section we describe the imprimitive non-monomial subgroups of $ \C_2 $. We reference the groups we find against the group libraries in GAP.

In the appendix we describe the other primitive subgroups of $\SU(4)$. These are either exotic (meaning they are outside of the Clifford hierarchy) or they are Clifford subgroups that do not contain the Pauli group. We also generalize the imprimitive subgroups of $ \C_2 $ to construct many groups appearing in the higher levels of the two-qubit Clifford hierarchy.

\subsection{Notation}

Since global phase is unphysical, the number of quantum operations in a given finite subgroup $ \G $ of $ \SU(4) $ is really the projective order of $ \G $ in $ \PU(4):=\SU(4)/i\I\I $. To emphasize this, we will follow the notation of  \cite{SU3Fairbairn1964FiniteAD}, where a finite matrix group is denoted by its projective order followed by a symbol indicating the order of its lift from the projective group (the total order is the product of the projective order and the order of the lift). 

Every subgroup of $ \PU(4) $ has at least an order $ 4 $ lift (which we denote with the symbol $ \sigma $) but not all subgroups have order $ 2 $ lifts (which we denote with the symbol $ \tau $) and only one of the groups we look at has only an order $ 1 $ lift to $ \SU(4) $. 

To be clear, $ \sigma $ will denote a lift of order $ 4 $, meaning the center is generated by $ i\I\I $, while $ \tau $ denotes a lift of order $ 2 $, meaning the center is generated by $ -\I\I $, and the lack of any symbol means the group has an order $ 1 $ (faithful) lift from $ \PU(4) $ to $ \SU(4) $ (and so has trivial center). Since the projective order counts the number of distinct quantum operations the group corresponds to, this is the main number we will denote our groups by. The symbols $ \tau, \sigma $ are only used to keep track of the size of the group for a specific choice of matrix generators.

In some cases we find two groups with similar properties, including the same order, but which are not isomorphic. In these cases we often distinguish the groups by their natural character. The natural character of a matrix group is given by taking the trace of each matrix in the group. These trace values generate some subring of $ \mathbb{C} $ extending the integers $ \mathbb{Z} $. When the rings generated by the trace values differ, we say the natural characters of the two matrix groups are defined over different rings of algebraic integers. 

\section{Primitive Clifford}

\subsection{Local}

We will start by examining primitive groups of local Clifford type, i.e., groups composed of matrices $ \mathsf{U} \otimes \mathsf{V}$ where $\mathsf{U}$ and $\mathsf{V}$ are determinant-1 single-qubit Clifford gates. 

The largest group of this type is composed of all the local Clifford gates: 
\[
\C_2^\text{loc} := \C_1 \otimes \C_1. \numberthis
\]
Here $\otimes$ means to take the Kronecker (tensor) product of all elements of the 1st factor with all elements of the 2nd factor. This group can also be generated as $ \expval{ \S  \I, \H  \I, \I \S,\I  \H}$, where juxtaposition of gates means tensor product here and throughout the paper. This group has order $576\tau$ and is listed in \cite{subgroupsSU4} as Group-\rom{14}. It can be called in GAP as \texttt{SmallGroup(1152,157463)}.

The commutator subgroup of $\C_2^\text{loc}$ is
\[
{\C_2^\text{loc}}' = \C_1' \otimes \C_1'. \numberthis
\]
This group, which is listed in \cite{subgroupsSU4} as Group-\rom{10}, has order $ 144\tau $ and can be called in GAP as \texttt{SmallGroup(288,860)}. In addition to the Pauli group $\P_2$, this group contains the generators $\F\I$ and $\I \F$.  

The remaining two primitive groups of local Clifford gates are both of order $288\tau$. The first is
\[
     \C_1' \otimes \C_1, \numberthis
\]
containing the single-qubit Paulis and $\F$ in one factor and the single-qubit Cliffords in another factor. The group $\C_1 \otimes \C_1'$ is conjugate to $ \C_1' \otimes \C_1 $ via a $\Swap$ gate, so both groups are of the same type. This group is listed in \cite{subgroupsSU4} as Group-\rom{12} and can be called in GAP as \texttt{SmallGroup(576,8273)}.

The group of local Cliffords $\C_2^\text{loc}$ together with the $\Swap$ gate yields the non-entangling Clifford gates and will be denoted by $\C_2^{\bowtie}$. Of course $\C_2^{\bowtie}$ is no longer local, but its commutator subgroup is local and is the last of the four primitive groups of local Clifford type:
\[
  \C_2^{{\bowtie}'}   = \expval{ \S \H, \H \S, \F \F}. \numberthis
\]
The group $\C_2^{{\bowtie}'}$ has the same order $288\tau$ as $\C_1' \otimes \C_1$, but the two groups are not isomorphic. The group $\C_2^{{\bowtie}'} $ is listed in \cite{subgroupsSU4} as Group-\rom{11} and it can be called in GAP as \texttt{SmallGroup(576,8282)}.

\begin{table}[htp]
    \centering
    \begin{tabular}{c|c|c|c} \toprule
    Name & Order & Generators & \cite{subgroupsSU4}  \\ \midrule
    $\C_1 \otimes \C_1$ & $576\tau$ & $\expval{\S \I, \H \I,\I \S, \I \H}$ & \rom{14}  \\
    $\C_1' \otimes \C_1'$ & $144\tau$ & $\expval{\Z \I, \F \I,\I \Z, \I \F}$ & \rom{10}  \\
    $\C_1' \otimes \C_1$ & $288\tau$ & $\expval{\Z \I, \F \I,\I \S, \I \H }$ & \rom{12}  \\
    $\C_2^{{\bowtie}'}$ & $288\tau$ & $\expval{\S \H, \H \S, \F \F}$ & \rom{11}  \\ \bottomrule
    \end{tabular}
    \caption{Primitive Local Subgroups of $ \C_2$}
\end{table}

\subsection{Non-Entangling}

A gate is called non-entangling if it is a product of local gates and qubit permutations. As already stated, adding in a $\Swap$ gate to the local Clifford group $\C_2^\text{loc}$ yields the group of non-entangling Cliffords, which we denote by 
\[
\C_2^{\bowtie}= \C_1 \bowtie \C_1:= \expval*{\C_1 \otimes \C_1, \Swap}. \numberthis
\]
Here $ \C_1 \bowtie \C_1 $ denotes the group generated by $ \C_1 \otimes \C_1 $ and swapping the tensor factors. This is Group-\rom{21} in \cite{subgroupsSU4} and has order $1152\sigma$. 

% Although we cannot call this group directly in GAP (as it is too large), we can call the projective group of order $1152$ using \texttt{SmallGroup(1152,157849)}.

Similarly, adding a $\Swap$ gate to $ \C_1' \otimes \C_1'$ yields the group
\[
\C_1' \bowtie \C_1' := \expval*{\C_1' \otimes \C_1', \Swap}. \numberthis
\]
This group, which is Group-\rom{19} in \cite{subgroupsSU4}, has order $288\sigma$  and can be called in GAP as \texttt{SmallGroup(1152,155473)}.

Taking the symmetric generating set $ \S\H, \H\S, \F\F  $ for $ \C_2^{\bowtie'} $ and adding $ \Swap $ yields the group 
\[
\C(576\sigma)_{\ZZ[\zeta_8]}  := \expval{\S\H, \H\S, \F\F, \Swap}. \numberthis
\]
% Again this group is too large to be called in GAP, but the projective version of order 576 can be called using \texttt{SmallGroup(576,8654)}. 
This group has order $576\sigma$ and is Group-\rom{17} in \cite{subgroupsSU4}. The natural character of this group is defined over $ \ZZ[\zeta_8] $, which we use in a subscript to differentiate it from the next group whose natural character is defined over $ \ZZ[i] $. Taking the generating set $ \S\H, \H\S, \F\F  $ and adding the generator $\S \I \cdot \Swap$ yields another group 
\[
    \C(576\sigma)_{\ZZ[i]} := \expval{\S\H, \H\S, \F\F, \S \I \cdot \Swap}. \numberthis
\]
This group also has order $576\sigma$ but is not isomorphic to the previous group, and corresponds to Group-\rom{18} in \cite{subgroupsSU4}. Here $ \cdot $ denotes matrix multiplication. 

% The projective version of this group of order 576 can be called in GAP using \texttt{SmallGroup(576,8652)}.

\begin{table}[htp]
    \centering
    \begin{tabular}{c|c|c|c} \toprule
    Name & Order & Generators & \cite{subgroupsSU4}  \\ \midrule
$\C_1 \bowtie \C_1$ & $1152\sigma$ & $\expval{\S\I,\H\I, \Swap}$ & \rom{21}   \\
$\C_1' \bowtie \C_1'$ & $288\sigma$ & $\expval{\Z\I,\F\I, \Swap}$ & \rom{19}  \\
$\C(576\sigma)_{\ZZ[\zeta_8]}$ & $576\sigma$ & $\expval{\S \H, \H \S, \F \F, \Swap}$ & \rom{17
} \\
$\C(576\sigma)_{\ZZ[i]}$ & $576\sigma$ & $\expval{\S \H, \H \S, \F \F, \S \I \cdot \Swap}$ & \rom{18} \\
\bottomrule
    \end{tabular}
    \caption{Primitive Non-Entangling Subgroups of $ \C_2 $}
\end{table}

\subsection{Entangling}

We now turn to the subgroups of $ \C_2 $ containing $ \P_2 $ that are primitive and include entangling gates. All $9$ of these groups can be generated using the entangling gate $\BELL$.

The first (and largest) group is the two-qubit Clifford group $\C_2$ itself, which has order $11520\sigma$ and is listed in \cite{subgroupsSU4} as Group-\rom{30}. The group $\C_2$ contains the non-entangling Cliffords as well as the entangling gates $\CNOT_{12}$, $\CNOT_{21}$, $\BELL$. The two-qubit Clifford group is a maximal subgroup of $ \SU(4) $ and a unitary $ 3 $-design (the $ n $-qubit Clifford group is always a unitary $ 3 $-design for any $ n $ \cite{CliffordGroup3Design}).

% Note that we can obtain the Clifford group $\C_2$ by adding $\CNOT$ to any of the four local Clifford type groups or any of the four non-entangling Clifford type groups. 

The largest subgroup of $\C_2$ is the commutator subgroup $ \C_2' $ which has order $ 5760\sigma $ and can be generated as
\[
    \C_2' = \expval{\C_1' \otimes \C_1', \BELL}. \numberthis
\]
The group $\C_2' $ is called Group-\rom{29} in \cite{subgroupsSU4} and is also a unitary $ 3 $-design \cite{UnitarytGroups}. It is a perfect group and can be called in GAP as \texttt{PerfectGroup(23040,2)}. Also $ \C_2' $ is the commutator subgroup of the complex reflection group with Shephard-Todd number 31 \cite{UnitarytGroups}.

The next largest subgroup is 
\[
\C(1920\sigma)_{\ZZ[\zeta_8]}:= \expval{\P_2,  \Swap, \BELL}, \numberthis
\]
which has order $ 1920\sigma$ and is listed in \cite{subgroupsSU4} as Group-\rom{28}. We label it by $ \ZZ[\zeta_8] $, the ring of algebraic integers that its natural character is defined over, to differentiate it from other groups of the same order. 

% This group is too large for the SmallGroup library in GAP but its projective version of order 1920 can be called using \texttt{SmallGroup(1920, 240993)}.

% note that 4dit clock and shift Pauli is irreducible but does not contained regular two qubit Pauli. Also 4dit clock and shift Clifford does contain regular two qubit Pauli but is not contained in regular two qubit Clifford, the intersection of 4dit clock Clifford, order 3072, w regular 2 qubit Clifford, order 46080, is order 512.

The commutator subgroup of $ \C(1920\sigma)_{\ZZ[\zeta_8]} $ has a remarkable form; it is the Clifford group for a single Galois qudit \cite{UnitaryDesigns2007}\footnote{This qudit Clifford group is not the one generated by the clock and shift matrices. } of dimension $ 4 $:
\[
\C_1(\mathbb{F}_4)=\C(1920\sigma)_{\ZZ[\zeta_8]} '=\expval{\P_2, \H \H, \BELL}. \numberthis
\]
This group has order $ 960\sigma $ and is listed in \cite{subgroupsSU4} as Group-\rom{26}. It is a perfect group and can be called in GAP as \texttt{PerfectGroup(3840,2)}. The Clifford group for any number of Galois qudits is always a unitary $ 2 $-design \cite{UnitaryDesigns2007} so $ \C_1(\mathbb{F}_4) $, and the group $ \C(1920\sigma)_{\ZZ[\zeta_8]} $ containing it, are both unitary $ 2 $-designs. Also $ \C_1(\mathbb{F}_4) $ is the commutator subgroup of the complex reflection group with Shephard-Todd number 29 \cite{UnitarytGroups}.

These first four groups are unitary $ 2 $-designs and so are not contained in any positive dimensional subgroups of $ \SU(4) $. However the next five groups are contained in the lift to $ \SU(4) $ of the large (dimension $ 10 $, whereas $ \SU(4) $ is dimension $ 15 $) $ \SO(5) $ subgroup of $ \PU(4) $. This subgroup of $ \SU(4) $ is generated by $ i\I \I $ and the symplectic subgroup $ \Sp(2) \subset \SU(4) $ and so can be described as having the structure $ \Sp(2).2 $.

The largest of these next five groups is 
\[
\C(1920\sigma)_{\ZZ[i,\sqrt{2}]}:= \expval{\C_1 \otimes \P_1, \BELL}, \numberthis
\]
which has order $ 1920\sigma$ and is listed in \cite{subgroupsSU4} as Group-\rom{27}. 

% This group is too large for the SmallGroup library but the projective version of order 1920 can be called in GAP by \texttt{SmallGroup(1920, 240996)}.

The commutator subgroup of $ \C(1920\sigma)_{\ZZ[i,\sqrt{2}]} $ is 
\[
\C(960\tau)=\C(1920\sigma)_{\ZZ[i,\sqrt{2}]}'=\expval{\F \I,\BELL}, \numberthis
\]
which has order $ 960\tau$ and is listed in \cite{subgroupsSU4} as Group-\rom{25}. This group is perfect and can be called in GAP as \texttt{PerfectGroup(1920,6)}. Adding the generator $ i\I\I $ yields the group $ \C(960\sigma)= \expval{\C_1' \otimes \P_1, \BELL} $. 

 % The projective version of order 960 can be called in GAP using \texttt{SmallGroup( 960, 11358)}.

The next two groups are 
\[
\C(320\sigma):= \expval{ \P_2, \K, \BELL}, \numberthis
\]
which has order $ 320 \sigma $ (listed in \cite{subgroupsSU4} as Group-\rom{24}) and can be called in GAP as \texttt{SmallGroup(1280,1116380)}. And 
\[
\C(160\sigma):= \expval{ \P_2, \K^2, \BELL} \numberthis
\]
which has order $ 160 \sigma $, is listed as Group-\rom{23} in \cite{subgroupsSU4}, and can be called as \texttt{SmallGroup(640,21464)} in GAP .
Both the groups just described use the gate 
\[
    \mathsf{K} := \F\I \cdot \exp( i \tfrac{\pi}{4} Y\otimes Z) = \tfrac{1}{\sqrt{2}} \smqty( 1 & 0 & -i & 0 \\ 0 & -i & 0 & -1 \\ -i & 0 & 1 & 0 \\ 0 & 1 & 0 & i). \numberthis
\]

The last and smallest primitive subgroup of $ \C_2 $  containing $ \P_2 $ is 
\[
\C(80\sigma):= \expval{ \P_2, \BELL} \numberthis
\]
which has order $ 80\sigma $, is listed in \cite{subgroupsSU4} as Group-\rom{22}, and can be called in GAP as \texttt{SmallGroup(320,1586)}.

\begin{table}[htp]
    \centering
    \begin{tabular}{c|c|c|c} \toprule
    Name & Order & Generators & \cite{subgroupsSU4}  \\ \midrule
$\C_2$ & $11520\sigma$ & $\expval{\C_1 \otimes \C_1, \BELL}$ & \rom{30}   \\
$\C_2'$ & $5760\sigma$ & $\expval{\C_1' \otimes \C_1', \BELL}$ & \rom{29} \\
$\C(1920\sigma)_{\ZZ[\zeta_8]}$ & $1920\sigma$ & $\expval{\P_2,  \Swap, \BELL}$ & \rom{28} \\ 
$\C_1(\mathbb{F}_4)$ & $960\sigma$  & $\expval{\P_2, \H \H, \BELL}$ &\rom{26} \\ \midrule 
$\C(1920\sigma)_{\ZZ[i,\sqrt{2}]}$ & $1920\sigma$ & $\expval{\C_1 \otimes \P_1, \BELL}$ & \rom{27} \\
$\C(960\sigma)$ & $960\sigma$ & $\expval{\C_1' \otimes \P_1, \BELL}$ & \rom{25} \\
$\C(320\sigma)$ & $320 \sigma$ & $\expval{\P_2,  \mathsf{K}, \BELL}$ & \rom{24} \\
$\C(160\sigma)$ & $160 \sigma$ & $\expval{\P_2,  \mathsf{K}^2, \BELL }$ & \rom{23}\\
$\C(80\sigma)$ & $80\sigma$ & $\expval{\P_2, \BELL}$ & \rom{22} \\

\bottomrule
    \end{tabular}
    \caption{Primitive Entangling Subgroups of $\C_2$ (containing $\P_2$)}
\end{table}

\section{Imprimitive Monomial Clifford}

Let $\G$ be a group of monomial matrices.
The subgroup of diagonal matrices $\Delta$ in this case forms a normal subgroup. For two-qubit gates, the quotient $\G/\Delta$ of a group of monomial matrices by its diagonal subgroup will always be either the symmetric group $ S_4 $, the alternating group $A_4$, the dihedral group $D_4$, or the Klein-4 group $V_4$. We will call these permutation groups the ``shape" of $\G$. The manner in which monomial subgroups of $\SU(4)$ are described in \cite{subgroupsSU4} is rather misleading - a given monomial group is usually not generated by a diagonal group and one of the 4 shapes, but rather, we can only say the quotient is one of these 4 shapes.

\subsection{Monomial Clifford subgroups of shape $ S_4 $}

We begin with the $ 6 $ monomial Clifford subgroups $ \G $ such that $ \G/\Delta $ is the symmetric group $ S_4 $.

There is one group of shape $ S_4 $ and order $ 768 \sigma $ given by
\[
\mathsf{M}(768\sigma, S_4) := \expval{\Q^{(2)} \otimes \Q^{(2)}, \CNOT_{12}, \CNOT_{21}  }. \numberthis
\]
The diagonal subgroup of this group is $ \Delta = \expval{ \S \I, \I \S, \CZ }$  which has order $ 128$.

There are two groups of shape $ S_4 $ and order $ 384 \sigma $. The first is
\[
\M(384\sigma,S_4)_{\mathbb{Z}[i]} := \expval{\P_2,\CNOT_{12}\cdot \S\I, \CNOT_{21}\cdot \S\I, }, \numberthis
\]
% Alternate generators
% \M(384\sigma,S_4)_{\mathbb{Z}[i]} := \expval{\P_2,  \DCNOT, C\X, \S\S }
which can be called  as \texttt{SmallGroup(1536,408569063)} in GAP. As before, the group subscript is used to denote the ring that the natural character of this group is defined over, namely the Gaussian integers $\mathbb{Z}[i]$, and is used to distinguish this group from the other monomial group of order $384\sigma$ with shape $S_4$. The diagonal subgroup here is $\Delta = \expval{\Z \I, \I \Z, \S\S, \CZ\cdot \S\I }$ of order $64$.

The other monomial group of shape $S_4$ and order $ 384\sigma$ is  
\[
\M(384\sigma,S_4)_{\mathbb{Z}[i,\sqrt{2}]} := \expval{\P_2 , \CNOT_{12}, \CNOT_{21}, \S\S  }, \numberthis
\]
% Could also use C\Z or \CZ\cdot \S\I in place of \S\S 
which can be called as \texttt{SmallGroup(1536,408569058)} in GAP. The diagonal subgroup here is also $\Delta = \expval{\Z \I, \I \Z, \S\S, \CZ \cdot \S \I  }$ of order $64$. The natural character of this group is defined over the ring $\mathbb{Z}[i,\sqrt{2}]$. 

There is one group of shape $ S_4 $ and order $ 192 \sigma $ given by
\[
\M(192\sigma,S_4) := \expval{\P_2,\CNOT_{12}, \CNOT_{21}, \CZ}, \numberthis 
\]
which can be called as \texttt{SmallGroup(768,1085977)} in GAP. The diagonal subgroup here is $\Delta = \expval{\Z \I, \I \Z, \mathsf{CZ} }$ of order 32. 

There are two groups of shape $ S_4 $ and order $ 96 \sigma $. The first looks almost exactly like $ \M(384\sigma,S_4)_{\mathbb{Z}[i]} $ but with a slight difference in the third generator
\[
\M(96\sigma,S_4)_{\mathbb{Z}[i]} := \expval{\P_2,\CNOT_{12}\cdot \S\I,\CNOT_{21} \cdot \I\S  }, \numberthis
\]
% Alternate generators:
% \M(96\sigma,S_4)_{\mathbb{Z}[i]} := \expval{\P_2 ,\DCNOT, C\Y} 
and can be called as \texttt{SmallGroup(384,20096)} in GAP. The diagonal subgroup here is $\Delta = \expval{i\I\I, \Z \I, \I \Z}$ of order $16$. The natural character of this group is defined over the Gaussian integers $\mathbb{Z}[i]$.

The other group of shape $S_4$ and order $ 96\sigma $ is 
\[
   \M(96\sigma,S_4)_{\mathbb{Z}[i,2 \zeta_8]}  := \expval{ \P_2, \CNOT_{12}, \CNOT_{21} }, \numberthis 
\]
which can be called as \texttt{SmallGroup(384,18142)} in GAP. The diagonal subgroup here is $\Delta = \expval{i\I\I, \Z \I, \I \Z}$ of order $16$. The natural character of this group is defined over $\mathbb{Z}[i,2\zeta_8]$.

\begin{table}[htp]
    \centering
    \begin{tabular}{c|c|c} \toprule
        Name  & Generators & Order \\ \midrule 
         $\M(768\sigma,S_4)$ & $\expval{\Q^{(2)} \otimes \Q^{(2)}, \CNOT_{12}, \CNOT_{21}  }$ & $768\sigma$ \\
         
         $\M(384\sigma,S_4)_{\mathbb{Z}[i]}$ & $\expval{\P_2, \CNOT_{12} \cdot \S\I, \CNOT_{21} \cdot \S\I }$ & $384\sigma$  \\

    $\M(384\sigma,S_4)_{\mathbb{Z}[i,\sqrt{2}]}$ & $\expval{\P_2 , \CNOT_{12}, \CNOT_{21} , \S\S  }$ & $384\sigma$  \\

    $\M(192\sigma,S_4)$ & $\expval{\P_2,\CNOT_{12}, \CNOT_{21}, \CZ}$ & $192\sigma$ \qquad  \\
    
    $\M(96\sigma,S_4)_{\mathbb{Z}[i]}$ & $\expval{\P_2 ,\CNOT_{12} \cdot \S\I, \CNOT_{21} \cdot \I\S }$ & $96\sigma$  \\
    
         $\M(96\sigma,S_4)_{\mathbb{Z}[i,2\zeta_8]}$ & $\expval{ \P_2, \CNOT_{12}, \CNOT_{21} }$ & $96\sigma$  \\ \bottomrule 
    \end{tabular}
    \caption{Imprimitive Subgroups of $ \C_2 $ that contain $ \P_2 $ and are Monomial of Shape $S_4$}
\end{table}

\subsection{Monomial Clifford subgroups of shape $ A_4 $}

Now we list the $ 4 $ monomial Clifford subgroups $ \G $ such that $ \G/\Delta $ is the alternating group $ A_4 $.

There is one group of shape $ A_4 $ and order $ 384 \sigma $ given by
\[
 \M(384\sigma,A_4) := \expval{\Q^{(2)} \otimes \Q^{(2)}, \DCNOT }. \numberthis
\]
This group is called \texttt{SmallGroup(1536,408535094)}  in GAP. The diagonal subgroup is $ \Delta = \expval{\S\I,\I\S, \CZ}$ and has order $ 128$.

The group of shape $ A_4 $ and order $ 192 \sigma $ is
\[
\M(192\sigma,A_4) := \expval{\P_2, \DCNOT, \S\S  } \numberthis.
\]
This group is called \texttt{SmallGroup(768,1083945)} in GAP. The diagonal subgroup is $\Delta = \expval{\Z\I,\I\Z,\S\S,\CZ\cdot \S\I }$ of order $64$.

The group of shape $ A_4 $ and order $ 96 \sigma $ is
\[
\M(96\sigma,A_4) := \expval{\P_2 ,\DCNOT, \CZ}, \numberthis
\]
which can be called as \texttt{SmallGroup(384,603)} in GAP. The diagonal subgroup is $\Delta = \expval{\Z\I,\I\Z, \CZ}$ of order 32. 

Finally, the group of shape $ A_4 $ of order $ 48 \sigma $ is
\[
\M(48\sigma,A_4) := \expval{\P_2, \DCNOT}, \numberthis
\]
which can be called as \texttt{SmallGroup(192,1509)} in GAP. The diagonal subgroup is $\Delta = \expval{i\I\I,\Z\I,\I\Z} $ of order $16$.

\begin{table}[htp]
    \centering
    \begin{tabular}{c|c|c} \toprule
        Name & Generators & Order  \\ \midrule 
        $\M(384\sigma,A_4)$  & $\expval{\Q^{(2)} \otimes \Q^{(2)}, \DCNOT }$ & $384\sigma$  \\
        
         $\M(192\sigma,A_4)$ & $\expval{\P_2, \DCNOT, \S\S  }$ & $192\sigma$  \\
         
         $\M(96\sigma,A_4)$ & $\expval{\P_2 ,\DCNOT, \CZ}$ & $96\sigma$ \\
         
         $\M(48\sigma,A_4)$ & $\expval{\P_2, \DCNOT}$ & $48\sigma$ \\
         \bottomrule
    \end{tabular}
    \caption{Imprimitive Subgroups of $ \C_2 $ that contain $ \P_2 $ and are Monomial of Shape $A_4$}
\end{table}

\subsection{Monomial Clifford subgroups of shape $ D_4 $}

Now we list the $ 13 $ monomial Clifford subgroups $ \G $ such that $ \G/\Delta $ is the dihedral group $ D_4 $.

There is one group of shape $ D_4 $ and order $ 256 \sigma $ given by
\[
 \M(256\sigma,D_4) := \expval{\Q^{(2)} \otimes \Q^{(2)}, \CNOT }. \numberthis
\]
The diagonal subgroup $ \Delta = \expval{\S\I, \I \S, \CZ} $ has order $ 128$.
% This group is unavailable in GAP due to it having order $1024$, but the projective version of this group can be called as \texttt{SmallGroup(256,26547)}.

There are $ 6 $ groups of shape $ D_4 $ of order $ 128 \sigma $. We have given these groups arbitrary labels of $a-f$ because there is no good way to distinguish them by character values alone. The first subgroup of order $128\sigma$ is
\[
\M(128\sigma,D_4)_a := \expval{\Q^{(2)} \otimes \P_1, \CNOT, \CZ }, \numberthis
\]
which can be called as \texttt{SmallGroup(512,419131)} in GAP. The diagonal subgroup is $\Delta = \expval{\S \I, \I\Z, \CZ} $ of order $64$. The second subgroup of order $128\sigma$ is
\[
\M(128\sigma,D_4)_b := \expval{\Q^{(2)} \otimes \P_1, \CNOT \cdot \I\S }, \numberthis
\]
which can be called in GAP as \texttt{SmallGroup(512,60109)}. The diagonal subgroup is $\Delta = \expval{\S \I, \I\Z, \CZ} $ of order $64$. The third subgroup of order $128\sigma$ is 
\[
\M(128\sigma,D_4)_c := \expval{\P_1 \otimes \Q^{(2)},\CNOT }, \numberthis 
\]
which can be called in GAP as \texttt{SmallGroup(512,59383)}. The diagonal subgroup is $\Delta = \expval{\Z\I, \I\S, \CZ \cdot \S\I}$ of order $64$. The fourth subgroup of order $128\sigma$ is
\[
\M(128\sigma,D_4)_d := \expval{\P_1 \otimes \Q^{(2)}, \CNOT \cdot \S\I, \CZ \cdot \S\I  }, \numberthis
\]
which can be called as  \texttt{SmallGroup(512,420089)} in GAP. The diagonal subgroup is $\Delta = \expval{\Z\I, \I\S, \CZ \cdot \S\I}$ of order $64$. The fifth subgroup of order $128\sigma$ is
\[
\M(128\sigma,D_4)_e := \expval{\P_2,\CNOT, \S\S }, \numberthis
\]
which can be called in GAP as \texttt{SmallGroup(512,60476)}. The diagonal subgroup is $\Delta = \expval{\Z\I, \I\Z, \S\S, \CZ \cdot \S\I}$ of order $64$. The sixth subgroup of order $128\sigma$ is
\[
\M(128\sigma,D_4)_f := \expval{\P_2, \CNOT \cdot \S\I , \S\S }, \numberthis
\]
which can be called in GAP as \texttt{SmallGroup(512,60321)}. The diagonal subgroup is $\Delta = \expval{\Z\I, \I\Z, \S\S, \CZ \cdot \S\I}$ of order $64$.

Next there are $ 5 $ groups of shape $ D_4 $ of order $ 64 \sigma $. Again we use the arbitrary labels $a-e$ for these groups because there is no good way to distinguish them by character values alone. The first subgroup of order $64\sigma$ is
\[
\M(64\sigma,D_4)_a := \expval{ \Q^{(2)} \otimes \P_1, \CNOT }, \numberthis
\]
which can be called in GAP by \texttt{SmallGroup(256,17275)}. The diagonal subgroup is $\Delta = \expval{i\I\I, \S\I, \I\Z}$ of order $32$. The second subgroup of order $64\sigma$ is
\[
\M(64\sigma,D_4)_b := \expval{\P_2 ,\CNOT,  \CZ}, \numberthis 
\]
which can be called in GAP by \texttt{SmallGroup(256,6039)}. The diagonal subgroup is $\Delta = \expval{\Z\I, \I\Z, \CZ}$ of order $32$. The third subgroup of order $64\sigma$ is
\[
\M(64\sigma,D_4)_c := \expval{\P_2, \CNOT \cdot \I \S,  \CZ \cdot \S \I }, \numberthis
\]
which can be called in GAP by \texttt{SmallGroup(256,6552)}. The diagonal subgroup is $\Delta = \expval{\Z\I,\I\Z,\CZ\cdot \S\I }$ of order $32$. The fourth subgroup of order $64\sigma$ is
\[
\M(64\sigma,D_4)_d := \expval{\P_2, \CNOT \cdot \S \S,  \CZ \cdot \S\I }, \numberthis 
\]
which can be called in GAP by \texttt{SmallGroup(256,6560)}. The diagonal subgroup is $\Delta = \expval{\Z\I,\I\Z,\CZ\cdot \S\I }$ of order $32$. The fifth subgroup of order $64\sigma$ is
\[
\M(64\sigma,D_4)_e := \expval{\P_2, \CNOT \cdot \S\I, \CZ \cdot \S\I  }, \numberthis
\]
which can be called in GAP by \texttt{SmallGroup(256,26555)}. The diagonal subgroup is $\Delta = \expval{\Z\I,\I\Z,\CZ\cdot \S\I }$ of order $32$.

Finally the group of shape $ D_4 $ of order $ 32\sigma $ is
\[
\M(32\sigma,D_4) := \expval{\P_2, \CNOT }, \numberthis
\]
which can be called in GAP by \texttt{SmallGroup(128,523)}. The diagonal subgroup is $ \Delta=\expval{i\I\I,\I\Z,\Z \I} $ of order $ 16$.

\begin{table}[htp]
    \centering
    \begin{tabular}{c|c|c} \toprule
        Name & Generators & Order  \\ \midrule 
    $\M(256\sigma,D_4)$ & $\expval{\Q^{(2)} \otimes \Q^{(2)}, \CNOT }$ & $256\sigma$ \\
    
    $\M(128\sigma,D_4)_a$ & $\expval{\Q^{(2)} \otimes \P_1, \CNOT, \CZ }$  & $128\sigma$ \\
    
    $\M(128\sigma,D_4)_b$ & $\expval{\Q^{(2)} \otimes \P_1, \CNOT \cdot \I\S }$ & $128\sigma$ \\
    
    $\M(128\sigma,D_4)_c$ & $\expval{\P_1 \otimes \Q^{(2)},\CNOT }$ & $128\sigma$ \\
    
    $\M(128\sigma,D_4)_d$ & $ \expval{\P_1 \otimes \Q^{(2)}, \CNOT \cdot \S\I, \CZ \cdot \S\I  }$  & $128\sigma$ \\
    
    $\M(128\sigma,D_4)_e$ & $\expval{\P_2,\CNOT, \S\S }$ & $128\sigma$ \\
    
    $\M(128\sigma,D_4)_f$ & $\expval{\P_2, \CNOT \cdot \S\I , \S\S }$  & $128\sigma$ \\ 

    $\M(64\sigma,D_4)_a$ & $\expval{ \Q^{(2)} \otimes \P_1, \CNOT }$  & $64\sigma$ \\
    
    $\M(64\sigma,D_4)_b$ & $\expval{\P_2 ,\CNOT, \CZ}$  & $64\sigma$  \\
    
    $\M(64\sigma,D_4)_c$ & $\expval{\P_2, \CNOT \cdot \I \S,  \CZ \cdot \S \I }$  & $64\sigma$ \\
    
    $\M(64\sigma,D_4)_d$ & $\expval{\P_2, \CNOT \cdot \S \S,  \CZ \cdot \S\I }$ & $64\sigma$ \\
    
    $\M(64\sigma,D_4)_e$ & $ \expval{\P_2, \CNOT \cdot \S\I, \CZ \cdot \S\I  }$ & $64\sigma$ \\

    $\M(32\sigma,D_4)$ & $\expval{\P_2, \CNOT }$ & $32\sigma$ \\ \bottomrule
    \end{tabular}
    \caption{Imprimitive Subgroups of $ \C_2 $ that contain $ \P_2 $ and are Monomial of Shape $D_4$ }
\end{table}

\subsection{Monomial Clifford subgroups of shape $ V_4 $}

Now we list the $ 6 $ monomial Clifford subgroups $ G $ such that $ G/\Delta $ is $ V_4 $.

There is one group of shape $ V_4 $ and order $ 128 \sigma $ given by
\[
 \M(128\sigma,V_4) := \expval{\Q^{(2)} \otimes \Q^{(2)}, \CZ }. \numberthis
\]
This group can be called in GAP as \texttt{SmallGroup(512,7521281)}. 
The diagonal subgroup is $ \Delta = \expval{\S\I,\I\S,\CZ}$ of order $ 128$. 

The next group with shape $V_4$ is the local group of order $64\tau$ given by
\[
  \Q^{(2)} \otimes \Q^{(2)}. \numberthis
\]
which can be called in GAP by \texttt{SmallGroup(128,2024)}.
The diagonal subgroup is $ \Delta= \expval{ \S \I, \I \S} $ of order $32$. 

A group of shape $V_4$ of order $64\sigma$ is given by
\[
 \M(64\sigma,V_4) := \expval{\P_2, \S\S, \CZ\cdot \S\I }. \numberthis 
\]
This group can be called in GAP as \texttt{SmallGroup(256,24064)}.
The diagonal subgroup is $ \Delta= \expval{ \Z \I, \I\Z,  \S \S, \CZ\cdot \S\I } $ of order $64$.

Next we have a local group of order $32\tau$ given by
\[
\Q^{(2)} \otimes \P_1. \numberthis
\]
which can be called in GAP by \texttt{SmallGroup(64,257)}. The diagonal subgroup is $ \Delta= \expval{ \S \I , \I \Z   } $ of order $16$.

The next group of shape $V_4$ has order $32\sigma$ and is given by
\[
 \M(32\sigma,V_4):= \expval{\P_2, \CZ\cdot \S\I }. \numberthis
\]
This group can be called in GAP by \texttt{SmallGroup(128,1750)}.
The diagonal subgroup is $ \Delta= \expval{ \Z \I, \I\Z, \CZ\cdot \S\I } $ of order $32$.

\begin{table}[htp]
    \centering
    \begin{tabular}{c|c|c} \toprule
        Name & Generators & Order  \\ \midrule 
    $\M(128\sigma,V_4)$ & $\expval{\Q^{(2)} \otimes \Q^{(2)}, \CZ }$ & $128\sigma$ \\

    $\M(64\sigma,V_4)$ & $ \expval{\P_2, \S\S, \CZ\cdot \S\I  }$ & $64\sigma$ \\

    $\M(32\sigma,V_4)$ & $ \expval{\P_2, \CZ\cdot \S\I } $ & $32\sigma$ \\

    $\Q^{(2)} \otimes \Q^{(2)}$ & $\expval{\X\I,\S\I,\I\X,\I\S }$ & $64\tau$ \\

    $\Q^{(2)} \otimes \P_1$ & $ \expval{\X\I, \S\I, \I\X,\I\Z } $ & $32\tau$ \\

    $\P_2$ & $ \expval{ \X\I,\Z\I, \I\X,\I\Z }$ & $16\tau$ \\ \bottomrule
    
    \end{tabular}
    \caption{Imprimitive Subgroups of $ \C_2 $ that contain $ \P_2 $ and are Monomial of Shape $V_4$ }
\end{table}

The final group of shape $ V_4 $ is the two-qubit Pauli group of order $ 16 \tau $ 
\[
\P_2 := \P_1 \otimes \P_1. \numberthis
\]
which can be called in GAP by \texttt{SmallGroup(32,49)}. The diagonal subgroup is $ \Delta=\expval{ \Z \I, \I \Z} $ of order $ 8 $.

\section{Imprimitive Non-Monomial Clifford}

Now that we have classified the subgroups of $ \C_2 $ containing $ \P_2 $ that are either primitive or 
 are imprimitive and monomial, we can, finally, turn to the case of non-monomial imprimitive subgroups.

\subsection{Local Groups}

There are $ 5 $ non-monomial imprimitive Clifford subgroups consisting of only local gates. The largest of these local groups has order $192\tau$ and is given by 
\[
\Q^{(2)} \otimes \C_1. \numberthis
\]
and can be called in GAP as \texttt{SmallGroup(384,18044)}.

The next local group has order $96\tau$, is given by
\[
\P_1 \otimes \C_1, \numberthis
\]
and can be called in GAP as \texttt{SmallGroup(192,1484)}. 

Another local group of the same order $96\tau$ is given by
\[
\Q^{(2)} \otimes \C_1' \numberthis
\]
and can be called in GAP as \texttt{SmallGroup(192,1017)}.

The smallest of these local groups is
\[
    \P_1 \otimes \C_1'. \numberthis
\]
This group has order $48\tau$ and can be called in GAP as \texttt{SmallGroup(96,201)}. 

A final, and less obvious, local group is 
\[
\N(96\tau) := \expval{ \P_1 \otimes \C_1', \S \S  } \numberthis
\]
which has order $ 96\tau $, and can be called in GAP as \texttt{SmallGroup(192,988)}.

\begin{table}[htp]
    \centering
    \begin{tabular}{c|c} \toprule
        Name  & Order  \\ \midrule 
        $\Q^{(2)} \otimes \C_1$  & $192\tau$ \\
         $\P_1 \otimes \C_1$ & $96\tau$ \\
         $\Q^{(2)} \otimes \C_1'$  & $96\tau$ \\
        $\P_1 \otimes \C_1'$ &  $48\tau$ \\
         $\N(96\tau)$ & $96\tau$ \\ 
         \bottomrule
    \end{tabular}
    \caption{Non-Monomial Imprimitive Local Subroups of $ \C_2 $ containing $ \P_2 $}
\end{table}

\subsection{Entangling Groups}

There are $ 5 $ non-monomial imprimitive Clifford subgroups with entangling gates. We start with 
\[
\N(768\sigma) := \expval{\Q^{(2)} \otimes \C_1, \CNOT},\numberthis
\]
of order $ 768 \sigma $.

% This group is unavailable in GAP because it has order $3072$ but the projective version of this group can be called using \texttt{SmallGroup(768,1088565)}. 

Then we have $3$ groups of order $1536 = 384\sigma$. We will use the labels $a-c$ to distinguish them. The first of these groups is
\[
\N(384\sigma)_a := \expval{\Q^{(2)} \otimes \C_1', \CNOT }, \numberthis
\]
of order $ 384 \sigma $
which can be called in GAP as \texttt{SmallGroup(1536,408528836)}. Next is
\[
\N(384\sigma)_b := \expval{\P_1 \otimes \C_1, \CZ\cdot \S\I }, \numberthis
\]
also of order $384\sigma$, which can be called in GAP as \texttt{SmallGroup(1536,408557064)}. The third group of order $384\sigma$ is
\[
\N(384\sigma)_c := \expval{ \P_1 \otimes \C_1', \S \S , \CZ \cdot \S\I } \numberthis
\]
and it can be called as \texttt{SmallGroup(1536,408546526)} in GAP.

The next group has order $ 192\sigma $
\[
\N(192\sigma) := \expval{\P_1 \otimes \C_1', \CZ\cdot \S\I } \numberthis
\]
 and can be called in GAP as \texttt{SmallGroup(768,1084090)}.

\begin{table}[htp]
    \centering
    \begin{tabular}{c|c|c} \toprule
        Name & Generators & Order  \\ \midrule 
        $\N(768\sigma)$ & $\expval{\Q^{(2)} \otimes \C_1, \CNOT}$ & $768\sigma$ \\
        
        $\N(384\sigma)_a$ & $\expval{\Q^{(2)} \otimes \C_1', \CNOT }$ & $384\sigma$ \\
        
        $\N(384\sigma)_b$ & $\expval{\P_1 \otimes \C_1, \CZ\cdot \S\I }$ & $384\sigma$ \\
        
        $\N(384\sigma)_c$  & $\expval{ \P_1 \otimes \C_1', \S \S , \CZ \cdot \S\I }$ & $384\sigma$ \\
        
        $\N(192\sigma)$ & $\expval{\P_1 \otimes \C_1', \CZ\cdot \S\I }$ & $192\sigma$ \\
        
       \bottomrule
    \end{tabular}
    \caption{Non-Monomial Imprimitive Entangling Subgroups of $ \C_2 $ containing $ \P_2 $}
\end{table}

\newpage 
\section{Conclusion}
\vspace{-0.25cm}
We have classified all subgroups of the two-qubit Clifford group $\C_2$ that contain the two-qubit Pauli group $\P_2$ and we have presented them in modern notation familiar to the quantum information community. We have listed their GAP calls when possible and we have also noted any other interesting properties the groups possess. These subgroups are the only Clifford subgroups that can be the transversal gate group of an $ [[n,2,d]] $ stabilizer code. For convenience, we provide in the appendix a table with all $56$ of these Clifford subgroups that includes GAP calls to the group, GAP calls to the projective group, generators, order, and our naming convention. 

\vspace{0.5cm}
In the appendix we also classify all of the primitive subgroups of $\SU(4)$, again using modern notation familiar to the quantum information community. We also list several infinite series of groups that are in higher levels of the two-qubit Clifford hierarchy. 

\section*{Acknowledgements}

We thank Jonas Anderson and Markus Heinrich for helpful conversations on \textit{quantum computing stack exchange}. This research was supported in part by the MathQuantum RTG through the NSF RTG grant DMS-2231533.

\onecolumngrid
\bibliography{biblio} 

\begin{thebibliography}{20}%
\makeatletter
\providecommand \@ifxundefined [1]{%
 \@ifx{#1\undefined}
}%
\providecommand \@ifnum [1]{%
 \ifnum #1\expandafter \@firstoftwo
 \else \expandafter \@secondoftwo
 \fi
}%
\providecommand \@ifx [1]{%
 \ifx #1\expandafter \@firstoftwo
 \else \expandafter \@secondoftwo
 \fi
}%
\providecommand \natexlab [1]{#1}%
\providecommand \enquote  [1]{``#1''}%
\providecommand \bibnamefont  [1]{#1}%
\providecommand \bibfnamefont [1]{#1}%
\providecommand \citenamefont [1]{#1}%
\providecommand \href@noop [0]{\@secondoftwo}%
\providecommand \href [0]{\begingroup \@sanitize@url \@href}%
\providecommand \@href[1]{\@@startlink{#1}\@@href}%
\providecommand \@@href[1]{\endgroup#1\@@endlink}%
\providecommand \@sanitize@url [0]{\catcode `\\12\catcode `\$12\catcode
  `\&12\catcode `\#12\catcode `\^12\catcode `\_12\catcode `\%12\relax}%
\providecommand \@@startlink[1]{}%
\providecommand \@@endlink[0]{}%
\providecommand \url  [0]{\begingroup\@sanitize@url \@url }%
\providecommand \@url [1]{\endgroup\@href {#1}{\urlprefix }}%
\providecommand \urlprefix  [0]{URL }%
\providecommand \Eprint [0]{\href }%
\providecommand \doibase [0]{https://doi.org/}%
\providecommand \selectlanguage [0]{\@gobble}%
\providecommand \bibinfo  [0]{\@secondoftwo}%
\providecommand \bibfield  [0]{\@secondoftwo}%
\providecommand \translation [1]{[#1]}%
\providecommand \BibitemOpen [0]{}%
\providecommand \bibitemStop [0]{}%
\providecommand \bibitemNoStop [0]{.\EOS\space}%
\providecommand \EOS [0]{\spacefactor3000\relax}%
\providecommand \BibitemShut  [1]{\csname bibitem#1\endcsname}%
\let\auto@bib@innerbib\@empty
%</preamble>
\bibitem [{\citenamefont {Eastin}\ and\ \citenamefont
  {Knill}(2009)}]{eastinKnill}%
  \BibitemOpen
  \bibfield  {author} {\bibinfo {author} {\bibfnamefont {B.}~\bibnamefont
  {Eastin}}\ and\ \bibinfo {author} {\bibfnamefont {E.}~\bibnamefont {Knill}},\
  }\bibfield  {title} {\bibinfo {title} {Restrictions on transversal encoded
  quantum gate sets},\ }\bibfield  {journal} {\bibinfo  {journal} {Physical
  Review Letters}\ }\textbf {\bibinfo {volume} {102}},\ \href
  {https://doi.org/10.1103/physrevlett.102.110502}
  {10.1103/physrevlett.102.110502} (\bibinfo {year} {2009})\BibitemShut
  {NoStop}%
\bibitem [{\citenamefont {Jochym-O'Connor}\ \emph {et~al.}(2018)\citenamefont
  {Jochym-O'Connor}, \citenamefont {Kubica},\ and\ \citenamefont
  {Yoder}}]{disjointnessStabilizerCodes}%
  \BibitemOpen
  \bibfield  {author} {\bibinfo {author} {\bibfnamefont {T.}~\bibnamefont
  {Jochym-O'Connor}}, \bibinfo {author} {\bibfnamefont {A.}~\bibnamefont
  {Kubica}},\ and\ \bibinfo {author} {\bibfnamefont {T.~J.}\ \bibnamefont
  {Yoder}},\ }\bibfield  {title} {\bibinfo {title} {Disjointness of stabilizer
  codes and limitations on fault-tolerant logical gates},\ }\bibfield
  {journal} {\bibinfo  {journal} {Physical Review X}\ }\textbf {\bibinfo
  {volume} {8}},\ \href {https://doi.org/10.1103/physrevx.8.021047}
  {10.1103/physrevx.8.021047} (\bibinfo {year} {2018})\BibitemShut {NoStop}%
\bibitem [{\citenamefont {Gottesman}(1997)}]{stabilizerThesis}%
  \BibitemOpen
  \bibfield  {author} {\bibinfo {author} {\bibfnamefont {D.}~\bibnamefont
  {Gottesman}},\ }\href@noop {} {\bibinfo {title} {Stabilizer codes and quantum
  error correction}} (\bibinfo {year} {1997}),\ \Eprint
  {https://arxiv.org/abs/quant-ph/9705052} {arXiv:quant-ph/9705052 [quant-ph]}
  \BibitemShut {NoStop}%
\bibitem [{\citenamefont {Blichfeldt}(1917)}]{blichfeldt1917finite}%
  \BibitemOpen
  \bibfield  {author} {\bibinfo {author} {\bibfnamefont {H.}~\bibnamefont
  {Blichfeldt}},\ }\href {https://books.google.com/books?id=_FY7AQAAIAAJ}
  {\emph {\bibinfo {title} {Finite Collineation Groups: With an Introduction to
  the Theory of Groups of Operators and Substitution Groups}}},\ Cornell
  University Library historical math monographs\ (\bibinfo  {publisher}
  {University of Chicago Press},\ \bibinfo {year} {1917})\BibitemShut {NoStop}%
\bibitem [{\citenamefont {Hanany}\ and\ \citenamefont
  {He}(2001)}]{subgroupsSU4}%
  \BibitemOpen
  \bibfield  {author} {\bibinfo {author} {\bibfnamefont {A.}~\bibnamefont
  {Hanany}}\ and\ \bibinfo {author} {\bibfnamefont {Y.-H.}\ \bibnamefont
  {He}},\ }\bibfield  {title} {\bibinfo {title} {A monograph on the
  classification of the discrete subgroups of {SU}(4)},\ }\href
  {https://doi.org/10.1088/1126-6708/2001/02/027} {\bibfield  {journal}
  {\bibinfo  {journal} {Journal of High Energy Physics}\ }\textbf {\bibinfo
  {volume} {2001}},\ \bibinfo {pages} {027} (\bibinfo {year}
  {2001})}\BibitemShut {NoStop}%
\bibitem [{Note1()}]{Note1}%
  \BibitemOpen
  \bibinfo {note} {See the Supplemental Material of \cite {us1} for a history
  of the Facet gate.}\BibitemShut {Stop}%
\bibitem [{\citenamefont {Kubischta}\ and\ \citenamefont
  {Teixeira}(2024{\natexlab{a}})}]{us2}%
  \BibitemOpen
  \bibfield  {author} {\bibinfo {author} {\bibfnamefont {E.}~\bibnamefont
  {Kubischta}}\ and\ \bibinfo {author} {\bibfnamefont {I.}~\bibnamefont
  {Teixeira}},\ }\href {https://arxiv.org/abs/2310.17652} {\bibinfo {title}
  {The not-so-secret fourth parameter of quantum codes}} (\bibinfo {year}
  {2024}{\natexlab{a}}),\ \Eprint {https://arxiv.org/abs/2310.17652}
  {arXiv:2310.17652 [quant-ph]} \BibitemShut {NoStop}%
\bibitem [{\citenamefont {Hill}\ and\ \citenamefont
  {Wootters}(1997)}]{BellGate1997}%
  \BibitemOpen
  \bibfield  {author} {\bibinfo {author} {\bibfnamefont {S.~A.}\ \bibnamefont
  {Hill}}\ and\ \bibinfo {author} {\bibfnamefont {W.~K.}\ \bibnamefont
  {Wootters}},\ }\bibfield  {title} {\bibinfo {title} {Entanglement of a pair
  of quantum bits},\ }\href {https://doi.org/10.1103/PhysRevLett.78.5022}
  {\bibfield  {journal} {\bibinfo  {journal} {Phys. Rev. Lett.}\ }\textbf
  {\bibinfo {volume} {78}},\ \bibinfo {pages} {5022} (\bibinfo {year}
  {1997})}\BibitemShut {NoStop}%
\bibitem [{\citenamefont {Vatan}\ and\ \citenamefont
  {Williams}(2004)}]{BellGate2004}%
  \BibitemOpen
  \bibfield  {author} {\bibinfo {author} {\bibfnamefont {F.}~\bibnamefont
  {Vatan}}\ and\ \bibinfo {author} {\bibfnamefont {C.}~\bibnamefont
  {Williams}},\ }\bibfield  {title} {\bibinfo {title} {Optimal quantum circuits
  for general two-qubit gates},\ }\bibfield  {journal} {\bibinfo  {journal}
  {Physical Review A}\ }\textbf {\bibinfo {volume} {69}},\ \href
  {https://doi.org/10.1103/physreva.69.032315} {10.1103/physreva.69.032315}
  (\bibinfo {year} {2004})\BibitemShut {NoStop}%
\bibitem [{\citenamefont {Zeng}\ \emph {et~al.}(2008)\citenamefont {Zeng},
  \citenamefont {Chen},\ and\ \citenamefont {Chuang}}]{cliffordhierarchy}%
  \BibitemOpen
  \bibfield  {author} {\bibinfo {author} {\bibfnamefont {B.}~\bibnamefont
  {Zeng}}, \bibinfo {author} {\bibfnamefont {X.}~\bibnamefont {Chen}},\ and\
  \bibinfo {author} {\bibfnamefont {I.~L.}\ \bibnamefont {Chuang}},\ }\bibfield
   {title} {\bibinfo {title} {Semi-clifford operations, structure of
  {${\mathcal{C}}_{k}$} hierarchy, and gate complexity for fault-tolerant
  quantum computation},\ }\href {https://doi.org/10.1103/PhysRevA.77.042313}
  {\bibfield  {journal} {\bibinfo  {journal} {Phys. Rev. A}\ }\textbf {\bibinfo
  {volume} {77}},\ \bibinfo {pages} {042313} (\bibinfo {year}
  {2008})}\BibitemShut {NoStop}%
\bibitem [{\citenamefont {Kubischta}\ and\ \citenamefont
  {Teixeira}(2023)}]{us1}%
  \BibitemOpen
  \bibfield  {author} {\bibinfo {author} {\bibfnamefont {E.}~\bibnamefont
  {Kubischta}}\ and\ \bibinfo {author} {\bibfnamefont {I.}~\bibnamefont
  {Teixeira}},\ }\bibfield  {title} {\bibinfo {title} {Family of quantum codes
  with exotic transversal gates},\ }\bibfield  {journal} {\bibinfo  {journal}
  {Physical Review Letters}\ }\textbf {\bibinfo {volume} {131}},\ \href
  {https://doi.org/10.1103/physrevlett.131.240601}
  {10.1103/physrevlett.131.240601} (\bibinfo {year} {2023})\BibitemShut
  {NoStop}%
\bibitem [{GAP()}]{GAP}%
  \BibitemOpen
  GAP,\ \href@noop {} {\bibinfo {title} {{GAP} {\textendash} {G}roups,
  {A}lgorithms, and {P}rogramming, {V}ersion 4.12.0}},\ \bibinfo {howpublished}
  {\href {https://www.gap-system.org} {\texttt{https://www.gap-system.org}}}
  (\bibinfo {year} {2018})\BibitemShut {NoStop}%
\bibitem [{\citenamefont {Fairbairn}\ \emph {et~al.}(1964)\citenamefont
  {Fairbairn}, \citenamefont {Fulton},\ and\ \citenamefont
  {Klink}}]{SU3Fairbairn1964FiniteAD}%
  \BibitemOpen
  \bibfield  {author} {\bibinfo {author} {\bibfnamefont {W.}~\bibnamefont
  {Fairbairn}}, \bibinfo {author} {\bibfnamefont {T.}~\bibnamefont {Fulton}},\
  and\ \bibinfo {author} {\bibfnamefont {W.~H.}\ \bibnamefont {Klink}},\
  }\bibfield  {title} {\bibinfo {title} {Finite and disconnected subgroups of
  {SU}3 and their application to the elementary‐particle spectrum},\ }\href
  {https://api.semanticscholar.org/CorpusID:123080515} {\bibfield  {journal}
  {\bibinfo  {journal} {Journal of Mathematical Physics}\ }\textbf {\bibinfo
  {volume} {5}},\ \bibinfo {pages} {1038} (\bibinfo {year} {1964})}\BibitemShut
  {NoStop}%
\bibitem [{\citenamefont {Webb}(2016)}]{CliffordGroup3Design}%
  \BibitemOpen
  \bibfield  {author} {\bibinfo {author} {\bibfnamefont {Z.}~\bibnamefont
  {Webb}},\ }\bibfield  {title} {\bibinfo {title} {The clifford group forms a
  unitary 3-design},\ }\href@noop {} {\bibfield  {journal} {\bibinfo  {journal}
  {Quantum Info. Comput.}\ }\textbf {\bibinfo {volume} {16}},\ \bibinfo {pages}
  {1379–1400} (\bibinfo {year} {2016})}\BibitemShut {NoStop}%
\bibitem [{\citenamefont {Bannai}\ \emph {et~al.}(2018)\citenamefont {Bannai},
  \citenamefont {Navarro}, \citenamefont {Rizo},\ and\ \citenamefont
  {Tiep}}]{UnitarytGroups}%
  \BibitemOpen
  \bibfield  {author} {\bibinfo {author} {\bibfnamefont {E.}~\bibnamefont
  {Bannai}}, \bibinfo {author} {\bibfnamefont {G.}~\bibnamefont {Navarro}},
  \bibinfo {author} {\bibfnamefont {N.}~\bibnamefont {Rizo}},\ and\ \bibinfo
  {author} {\bibfnamefont {P.~H.}\ \bibnamefont {Tiep}},\ }\href
  {https://arxiv.org/abs/1810.02507} {\bibinfo {title} {Unitary t-groups}}
  (\bibinfo {year} {2018}),\ \Eprint {https://arxiv.org/abs/1810.02507}
  {arXiv:1810.02507 [math.RT]} \BibitemShut {NoStop}%
\bibitem [{\citenamefont {Gross}\ \emph {et~al.}(2007)\citenamefont {Gross},
  \citenamefont {Audenaert},\ and\ \citenamefont
  {Eisert}}]{UnitaryDesigns2007}%
  \BibitemOpen
  \bibfield  {author} {\bibinfo {author} {\bibfnamefont {D.}~\bibnamefont
  {Gross}}, \bibinfo {author} {\bibfnamefont {K.}~\bibnamefont {Audenaert}},\
  and\ \bibinfo {author} {\bibfnamefont {J.}~\bibnamefont {Eisert}},\
  }\bibfield  {title} {\bibinfo {title} {Evenly distributed unitaries: On the
  structure of unitary designs},\ }\bibfield  {journal} {\bibinfo  {journal}
  {Journal of Mathematical Physics}\ }\textbf {\bibinfo {volume} {48}},\ \href
  {https://doi.org/10.1063/1.2716992} {10.1063/1.2716992} (\bibinfo {year}
  {2007})\BibitemShut {NoStop}%
\bibitem [{Note2()}]{Note2}%
  \BibitemOpen
  \bibinfo {note} {This qudit Clifford group is not the one generated by the
  clock and shift matrices.}\BibitemShut {Stop}%
\bibitem [{\citenamefont {Gross}(2021)}]{gross1}%
  \BibitemOpen
  \bibfield  {author} {\bibinfo {author} {\bibfnamefont {J.~A.}\ \bibnamefont
  {Gross}},\ }\bibfield  {title} {\bibinfo {title} {Designing codes around
  interactions: The case of a spin},\ }\bibfield  {journal} {\bibinfo
  {journal} {Physical Review Letters}\ }\textbf {\bibinfo {volume} {127}},\
  \href {https://doi.org/10.1103/physrevlett.127.010504}
  {10.1103/physrevlett.127.010504} (\bibinfo {year} {2021})\BibitemShut
  {NoStop}%
\bibitem [{\citenamefont {Kubischta}\ and\ \citenamefont
  {Teixeira}(2024{\natexlab{b}})}]{us3}%
  \BibitemOpen
  \bibfield  {author} {\bibinfo {author} {\bibfnamefont {E.}~\bibnamefont
  {Kubischta}}\ and\ \bibinfo {author} {\bibfnamefont {I.}~\bibnamefont
  {Teixeira}},\ }\bibfield  {title} {\bibinfo {title} {Quantum codes from
  twisted unitary $t$-groups},\ }\href
  {https://doi.org/10.1103/PhysRevLett.133.030602} {\bibfield  {journal}
  {\bibinfo  {journal} {Phys. Rev. Lett.}\ }\textbf {\bibinfo {volume} {133}},\
  \bibinfo {pages} {030602} (\bibinfo {year} {2024}{\natexlab{b}})}\BibitemShut
  {NoStop}%
\bibitem [{\citenamefont {Anderson}(2024)}]{Anderson_2024}%
  \BibitemOpen
  \bibfield  {author} {\bibinfo {author} {\bibfnamefont {J.~T.}\ \bibnamefont
  {Anderson}},\ }\bibfield  {title} {\bibinfo {title} {On groups in the qubit
  clifford hierarchy},\ }\href {https://doi.org/10.22331/q-2024-06-13-1370}
  {\bibfield  {journal} {\bibinfo  {journal} {Quantum}\ }\textbf {\bibinfo
  {volume} {8}},\ \bibinfo {pages} {1370} (\bibinfo {year} {2024})}\BibitemShut
  {NoStop}%
\end{thebibliography}%

\newpage 

\appendix
\onecolumngrid
\section{ Primitive Clifford Subgroups That Do Not Contain the Pauli Group}

In the single qubit case, every irreducible subgroup of the Clifford group $\C_1$ contains the Pauli group $\P_1$. However, in the two-qubit case this is no longer true - there are irreducible subgroups (in fact, primitive subgroups) of the two-qubit Clifford group $\C_2$ that do not contain the two-qubit Pauli group $\P_2$. Although a group that does not contain the Pauli group cannot be the transversal gate group of a stabilizer code, such a group may be interesting for other reasons, and so we list them below.

The largest primitive subgroup of $ \C_2 $ not containing $ \P_2 $ is
\[
\C(360\tau)=\expval{\I\F,\BELL}. \numberthis
\]
% This group is also the commutator subgroup of the exotic group $ \Ex(720\sigma) $ (which is described in the next appendix)
% Note that using the generator $ \F \I $ instead of $\I\F$ yields $ \C(960\tau) $, a completely different group.
This is listed in \cite{subgroupsSU4} as Group-\rom{3}.  The group $ \C(360\tau) $ has order $ 360 \tau $, is isomorphic to $ 2.\Alt_6 \cong \SL(2,9) $, and can be called in GAP as \texttt{PerfectGroup(720)}.

Another group is 
\[
\C(120\sigma)_{\ZZ[\zeta_8]}:=\expval{\Swap,\BELL}, \numberthis
\]
which has order $ 120 \sigma $ and is listed in \cite{subgroupsSU4} as Group-\rom{8}. This group can be called in GAP as \texttt{SmallGroup(480,217)}. 

% The projective version of this group is \texttt{SmallGroup(120,34)} in GAP. 

The commutator subgroup of the previous group, listed in \cite{subgroupsSU4} as Group-\rom{2}, is given by
\[
\C(60)=\C(120\sigma)_{\ZZ[\zeta_8]}'=\expval{ \CZ^\dagger \cdot \Swap,\BELL}. \numberthis
\]
This group has order $ 60 $, is isomorphic to $ \Alt_5 $, and can be called in GAP as \texttt{PerfectGroup(60)}. The $\tau$ version of this group is $\mathbb{Z}_2 \times \Alt_5$ and results by adding in a factor of $-\I\I$ and can be called in GAP by \texttt{SmallGroup(120,35)}. The $\sigma$ version of this group is $\mathbb{Z}_4 \times A_5$ and results from adding in a factor of $i\I\I$ and can be called in GAP by \texttt{SmallGroup(240,92)}.
% $$
% group 2=\expval{\Swap \cdot \BELL \cdot \Swap,\BELL}
% $$

The penultimate group is 
\[
\C(120\tau)_{\ZZ[i]}=\expval{\H\Z \cdot \BELL,\S\Z\cdot \BELL}, \numberthis
\]
which has order $ 120 \tau $, is listed in \cite{subgroupsSU4} as Group-\rom{7}, and is isomorphic to $ 2.\Sym_5\cong \SL(2,5).2 $ (there are two versions of this subgroup, this is the Clifford version, the exotic representation is given later). This group can be called in GAP by \texttt{SmallGroup(240,89)}. 

% The projective version of this group is \texttt{SmallGroup(120,34)}\texttt{SymmetricGroup(5)} in GAP. 

The final group is the commutator subgroup of the previous, listed in \cite{subgroupsSU4} as Group-\rom{1}, and given by 
\[
\C(120\tau)_{\ZZ[i]}':= \expval{\mathsf{A}, \CZ \cdot \S\F}, \numberthis
\]
where we have defined the monomial matrix
\[
    \mathsf{A} := \smqty( 0 & 0 & -1 & 0 \\ 0 & -i & 0 & 0 \\ 1 & 0 & 0 & 0 \\ 0 & 0 & 0 & i). \numberthis
\]
This group has order $ 60 \tau $, is isomorphic to $ \SL(2,5)\cong 2.\Alt_5 $, and can be called in GAP as \texttt{PerfectGroup(120)}.

% The projective version of this group is also $\Alt_5$ but this group is a different extension than $\C(60\tau)$ (which is simply a direct product of $\mathbb{Z}_2$ and $\Alt_5$). 

\begin{table}[htp]
    \centering
    \begin{tabular}{c|c|c|c} \toprule
    Name & Order & Generators & \cite{subgroupsSU4}  \\ \midrule
    $\C(360\tau)$ & $360\tau$ & $\expval{\I\F,\BELL}$  & \rom{3} \\
    $\C(120\sigma)_{\mathbb{Z}[\zeta_8]}$ & $120\sigma$ & $\expval{\Swap,\BELL}$ & \rom{8} \\
    $\C(120\sigma)_{\mathbb{Z}[\zeta_8]}'$ & $60$ & $\expval{\CZ^\dagger \cdot \Swap,\BELL}$ & \rom{2}  \\
    $\C(120\sigma)_{\mathbb{Z}[i]}$ & $120\sigma$ & $\expval{\H\Z \cdot \BELL, \S\Z \cdot \BELL}$  & \rom{7} \\
    $\C(120\sigma)_{\mathbb{Z}[i]}'$ & $60\tau$ & $\expval{\mathsf{A}, \CZ \cdot \S\F}$ & \rom{1} \\ \bottomrule
    \end{tabular}
    \caption{Primitive $\C_2$ Subgroups not containing $\P_2$}
\end{table}

\newpage 
\section{Exotic Primitive Subgroups of $ \SU(4) $}

For completeness, in this appendix we list the remaining 9 primitive subgroups of $\SU(4)$. Three of these are local, one is non-entangling, and five contain entangling gates. Each of these groups is exotic, meaning that they contain gates not only outside of the Clifford group $\C_2$, but outside of the entire two-qubit Clifford hierarchy. It is interesting that all of the primitive subgroups of $\SU(4)$ are either Clifford or exotic but never in some level of the Clifford hierarchy greater than $2$. 

Of the finite subgroups of $\SU(2)$, only the three groups $ \C_1',\C_1,\ico $ are primitive, and of these only the binary icosahedral group $\ico$ is exotic \cite{gross1,us1,us3}. Recall that $ \ico $ is generated by $\expval{\Z,\mathsf{\Phi}}$ where 
\[
\mathsf{\Phi} = \tfrac{1}{2} \smqty( \varphi + i \varphi^{-1} & 1 \\ \sm 1 & \varphi - i \varphi^{-1} ), \numberthis
\]
and $ \varphi=\frac{1+\sqrt{5}}{2} $ is the golden ratio. The group $ \ico $ will play a role below.

\subsection{Local or Non-entangling Exotic
}

The largest exotic primitive group of local gates is
\[
 \ico \otimes \ico. \numberthis
\]
where $ \ico$ is the binary icosahedral subgroup of $\SU(2)$. This group can be generated as $ \expval{ \Z  \I, \mathsf{\Phi}  \I, \I \Z,\I  \mathsf{\Phi}}$. This group has order $3600\tau$, is listed in \cite{subgroupsSU4} as Group-\rom{16}, and can be called in GAP as \texttt{PerfectGroup(7200,2)}. 

Another primitive local exotic group is
\[
 \ico \otimes \C_1. \numberthis
\]
This group can also be generated as $ \expval{ \Z  \I, \mathsf{\Phi}  \I, \I \S,\I  \H}$. This group has order $1440\tau$ and is listed in \cite{subgroupsSU4} as Group-\rom{15}. 

The last exotic primitive group of local gates is the commutator subgroup of the previous:
\[
\ico \otimes \C_1'. \numberthis
\]
This group can be generated as $ \expval{ \Z  \I, \mathsf{\Phi}  \I, \I \Z,\I  \F}$. This group has order $720\tau$ and is listed in \cite{subgroupsSU4} as Group-\rom{13} and can be called in GAP as \texttt{SmallGroup(1440,4615)}.

By adding $ \Swap $ to $\ico \otimes \ico $ we obtain the one non-entangling exotic primitive finite subgroup of $ \SU(4) $:
\[
 \ico \bowtie \ico. \numberthis
\]
This group has order $7200\sigma$ and is listed in \cite{subgroupsSU4} as Group-\rom{20}. 

\begin{table}[htp]
    \centering
    \begin{tabular}{c|c|c|c} \toprule
    Name & Order & Generators & \cite{subgroupsSU4}  \\ \midrule
    $\ico \otimes \ico$ & $3600\tau$ & $\expval{ \Z  \I, \mathsf{\Phi}  \I, \I \Z,\I  \mathsf{\Phi}}$ & \rom{16}  \\
    $\ico \otimes \C_1$ & $1440\tau$ & $\expval{ \Z  \I, \mathsf{\Phi}  \I, \I \S,\I  \H}$ & \rom{15}  \\
   $\ico \otimes \C_1'$ & $720\tau$ & $\expval{ \Z  \I, \mathsf{\Phi}  \I, \I \Z,\I  \F}$ & \rom{13}  \\
    \midrule 
    $\ico \bowtie \ico$ & $7200\sigma$ & $\expval{ \Z  \I, \mathsf{\Phi}  \I, \Swap}$ & \rom{20}  \\ \bottomrule
    \end{tabular}
    \caption{Primitive Exotic Groups (local and non-entangling)}
\end{table}

\subsection{Entangling Exotic}

We now turn to the primitive finite subgroups of $ \SU(4) $ that are exotic and contain entangling gates. We define $s = (1+ \sqrt{\sm 7})/2$ and $\overline{s} = (1-\sqrt{\sm 7})/2$. We will need the following matrices:
\begin{align*}
\mathsf{U}_1 := \tfrac{1}{\sqrt{ \sm 3}} \smqty( \sqrt{\sm3} & 0 & 0 & 0 \\ 0 & 1 & 1 & 1 \\ 0 & 1 & \zeta_3 & \zeta_3^* \\ 0 & 1 & \zeta_3^* & \zeta_3),\quad
  \mathsf{U}_2 &:= \smqty(0 & 0 & 1 & 0 \\ 0 & \sm1 & 0 & 0 \\ 1 & 0 & 0 & 0 \\ 0 & 0 & 0 & 1), \\
    \mathsf{V}_1 := \smqty(1 & 0 & 0 & 0 \\ 0 & 0 & 0 & \zeta_7^* \\ 0 & \zeta_7^5 & 0 & 0 \\ 0 & 0 & \zeta_7^3 & 0),\quad 
    \mathsf{V}_2 &:= \tfrac{1}{\sqrt{\sm 7}} \smqty(  \overline{s}^2 & 1 & 1 & 1 \\ 1 & s & \overline{s} & \overline{s} \\
1 & \overline{s} & s & \overline{s} \\ 1 & \overline{s} & \overline{s} & s), \\
    \mathsf{W}_1 := \tfrac{1}{\sqrt{3}}\smqty(1 & 0 & 0 & \sqrt{2} \\ 0 & \sm1 & \sqrt{2} & 0 \\ 0 & \sqrt{2} & 1 & 0 \\ \sqrt{2} & 0 & 0 & \sm 1 ),\quad
    \mathsf{W}_2 &:= \smqty( \sqrt{3}/2 & 1/2 & 0 & 0 \\ 1/2 & \sm \sqrt{3}/2 & 0 & 0 \\  0 & 0 & 0 & 1 \\ 0 & 0 & 1 & 0 ), \quad 
    \mathsf{W}_3 := \smqty(1 & 0 & 0 & 0 \\ 0 & 1 & 0 & 0 \\ 0 & 0 & \zeta_3 & 0 \\ 0 & 0 & 0 & \zeta_3^*)  
\end{align*}

The largest such group is 
\[
\Ex(25920\tau) := \expval{ \mathsf{U}_1, \mathsf{U}_2, \DCNOT}, \numberthis
\]
which has order $ 25920\tau $ and is listed in \cite{subgroupsSU4} as Group-\rom{6}. This group is isomorphic to the quasisimple finite group of Lie type $ \Sp(4,3) $ and can be called in GAP as $ \texttt{PerfectGroup(51840)}$. This group is defined over the ring $ \ZZ[\zeta_3] $, known as the Eisenstein integers. The group $\Ex(25920\tau) $ is the commutator subgroup of the complex reflection group with Shephard-Todd number 32 \cite{UnitarytGroups}. Also  $\Ex(25920\tau) $ is a unitary $ 3 $-design, and moreover a maximal subgroup of $ \SU(4) $.

The second largest group, listed in \cite{subgroupsSU4} as Group-\rom{4}, is 
\[
\Ex(2520\tau) := \expval{\mathsf{V}_1, \mathsf{V}_2, \DCNOT}, \numberthis
\]
which has order $ 2520\tau $ and can be called in GAP as $ \texttt{PerfectGroup(5040)}$. This group is isomorphic to the quasisimple finite group $ 2.\Alt_7 $ and its natural character is defined over the ring $ \ZZ[\tfrac{1+\sqrt{\sm 7}}{2}] $, known as the Kleinian integers. Also  $\Ex(2520\tau) $ is a unitary $ 3 $-design, and a maximal subgroup of $ \SU(4) $.

A maximal subgroup of $ \Ex(2520\tau) $ is
\[
\Ex(168\tau) := \expval{\mathsf{V}_1, \mathsf{V}_2}, \numberthis
\]
which has order $ 168\tau $ and is listed in \cite{subgroupsSU4} as Group-\rom{5}. This group is isomorphic to the quasisimple finite group $ 2.\GL(3,2)\cong\SL(2,7) $ and can be called in GAP as $ \texttt{PerfectGroup(336)}$. The group $ \Ex(168\tau) $ is contained in $ \Ex(2520\tau) $, so it is also defined over the Kleinian integers  $ \ZZ[\tfrac{1+\sqrt{\sm 7}}{2}] $. Although $ \Ex(168\tau) $ is contained in $ \Ex(2520\tau) $, it is not contained in any positive dimensional subgroup, a maximality property known as \textit{Lie primitivity}.

Next we consider the two subgroups 
$\Ex(720\tau)_{\ZZ[\sqrt{3}]}$ and $ \Ex(720\tau)_{\ZZ[\sqrt{\sm 3}]} $.
Both groups are isomorphic to the almost quasisimple group $ 2.\S_6 $. Adding $ \sigma=i \I \I$ to either group yields the same group 
\[
\Ex(720\sigma) := \expval{\mathsf{W}_1,\mathsf{W}_2, \mathsf{W}_3, \X\Z,\Z\X}, \numberthis
\]
which is listed as Group-\rom{9} in \cite{subgroupsSU4}.

\begin{table}[htp]
    \centering
    \begin{tabular}{c|c|c} \toprule
    Name & Generators   & \cite{subgroupsSU4}  \\ \midrule
    $\Ex(25920\tau)$ & $\expval{ \mathsf{U}_1, \mathsf{U}_2, \DCNOT}$ & \rom{6} \\
    $\Ex(2520\tau)$ & $\expval{\mathsf{V}_1, \mathsf{V}_2, \DCNOT}$ & \rom{4} \\
    $\Ex(168\tau)$ & $\expval{\mathsf{V}_1, \mathsf{V}_2}$ & \rom{5} \\
    $\Ex(720\sigma)$ & $\expval{\mathsf{W}_1,\mathsf{W}_2, \mathsf{W}_3, \X\Z,\Z\X}$  & \rom{9} \\
    $\Ex(120\sigma)$ & $\expval{\mathsf{W}_1,\mathsf{W}_2, \mathsf{W}_3, \X\Z \cdot \Z\X}$ & \rom{7} \\ \bottomrule
    \end{tabular}
    \caption{Primitive Exotic Groups (entangling)}
\end{table}

Finally we have the exotic version of the group $ 2.\Sym_5=\SL(2,5).2 $, listed as Group-\rom{7} in \cite{subgroupsSU4}, which corresponds to two inequivalent subgroups of $\SU(4) $. The Clifford version is defined over $\ZZ[i] $, which we have already listed as $ \C(120\tau)_{\ZZ[i]}$, and the exotic version is defined over $\ZZ[i,\sqrt{3}] $. We denote this group $\Ex(120\tau)$. If we add in $\sigma = i\I\I$ we get the group
\[
\Ex(120\sigma) := \expval{\mathsf{W}_1,\mathsf{W}_2, \mathsf{W}_3, \X\Z \cdot \Z\X}, \numberthis
\]
which can be called as \texttt{SmallGroup(480,946)} in GAP.

\section{Examples of Imprimitive Groups in Higher Levels of the Clifford Hierarchy}

In this appendix we will look at some irreducible groups in higher levels of the two-qubit Clifford hierarchy ($r \geq 3$). Note that all of these groups must be imprimitive since we have seen that all primitive groups in $\SU(4)$ are either Clifford subgroups or exotic.

\subsection{Monomial Groups in Higher Levels of the Clifford Hierarchy}

Many of the monomial imprimitive groups we have listed generalize to higher levels of the Clifford hierarchy in a natural way. We do not intend to be exhaustive and simply list a few examples, all of which fit into the paradigm described in \cite{Anderson_2024}.

\subsubsection{Shape $S_4$}
Recall that $\M(768\sigma,S_4)$ was generated by $\expval{\Q^{(2)} \otimes \Q^{(2)}, \CNOT_{12}, \CNOT_{21} }$. If we replace each $\Q^{(2)}$ by $\Q^{(r)}$ then we will have a monomial group, still of shape $S_4$, given by
\[
\M(24\cdot 2^{3r-1}\sigma , S_4) := \expval{\Q^{(r)} \otimes \Q^{(r)}, \CNOT_{12}, \CNOT_{21} }. \numberthis
\]
These groups are in the $r$-th level of the Clifford hierarchy for $r \geq 2$ (and the $r=1$ case is in the Clifford group). 

% Indeed the projective order of this group is $24 \cdot 2^{3r-1}$ (although the $r=1$ case is still in the 2nd level of the hierarchy). 

% Note that in this case if the factors are different, say $\Q^{(r_1)} \otimes \Q^{(r_2)}$ then you get the group $\M(24\cdot 2^{3\max(r_1,r_2)-1}\sigma , S_4)$.

\subsubsection{Shape $A_4$}
We also have a monomial series of shape $A_4$:
\[
\M(12\cdot 2^{3r-1}\sigma , A_4) := \expval{\Q^{(r)} \otimes \Q^{(r)}, \DCNOT }. \numberthis
\]
These groups are in the $r$-th level of the Clifford hierarchy for $r \geq 2$ (and the $r=1$ case is in the Clifford group). 

% Again if the factors are different, $\Q^{(r_1)} \otimes \Q^{(r_2)}$, then you get the group $\M(12\cdot 2^{3\max(r_1,r_2)-1}\sigma , A_4)$.

\subsubsection{Shape $D_4$}
We also have a monomial series of shape $D_4$:
\[
 \M(8\cdot 2^{3r-1}\sigma,D_4) := \expval{\Q^{(r)} \otimes \Q^{(r)}, \CNOT }. \numberthis
\]
These groups are in the $r$-th level of the Clifford hierarchy for $r \geq 2$ (and the $r=1$ case is in the Clifford group). However, these groups are not invariant under the $ \Swap $ gate and so if the factors are different we get a more general series:
\[
 \M(8\cdot 2^{(r_1+2r_2)-1}\sigma,D_4) := \expval{\Q^{(r_1)} \otimes \Q^{(r_2)}, \CNOT }. \numberthis
\]
Of course if $r_1 = r_2$ this reduces to the previous series. This series is in the $\max(r_1,r_2)$ level of the Clifford hierarchy (and the $r_1=r_2=1$ case is in the Clifford group). 

Now recall the $D_4$ groups $\M(128\sigma,D_4)_a = \expval{\Q^{(2)} \otimes \P_1, \CNOT, \CZ }$ and $\M(64\sigma,D_4)_b := \expval{\P_2 ,\CNOT,  \CZ}$. These are part of the same series:
\[
    \M(32 \cdot 2^r \sigma, D_4) := \expval{ \Q^{(r)} \otimes \P_1, \CNOT,\CZ}. \numberthis
\]
These groups are in the $r$-th level of the Clifford hierarchy for $r \geq 2$ (and the $r=1$ case is in the Clifford group). 

Next recall the $D_4$ groups $\M(128\sigma,D_4)_c = \expval{\P_1 \otimes \Q^{(2)},\CNOT }$ and $\M(32\sigma,D_4) = \expval{\P_2, \CNOT }$. These are part of the same series:
\[
\M(8\cdot 4^r\sigma,D_4) := \expval{\P_1 \otimes \Q^{(r)},\CNOT}. \numberthis
\]
These groups are in the $r$-th level of the Clifford hierarchy for $r \geq 2$ (and the $r=1$ case is in the Clifford group). 

Lastly, recall the $D_4$ groups $\M(64\sigma,D_4)_a = \expval{ \Q^{(2)} \otimes \P_1,\CNOT }$ and $\M(32\sigma,D_4) = \expval{\P_2, \CNOT }$. These are part of the same series:
\[
\M(16\cdot 2^r\sigma,D_4) := \expval{ \Q^{(r)} \otimes \P_1,\CNOT}. \numberthis
\]
These groups are in the $r$-th level of the Clifford hierarchy for $r \geq 2$ (and the $r=1$ case is in the Clifford group).

\subsubsection{Shape $V_4$}

Recall the group $\M(128\sigma,V_4) = \expval{\Q^{(2)} \otimes \Q^{(2)},\CZ}$. This is part of the series
\[
\M( 8 \cdot 2^{r_1+r_2}\sigma ,V_4) := \expval{ \Q^{(r_1)} \otimes \Q^{(r_2)}, \CZ  }, \numberthis
\]
for $r_1,r_2 \geq 1$. These groups are in the $\max(r_1,r_2)$ level of the Clifford hierarchy (and the $r_1=r_2=1$ case is in the Clifford group).

The $V_4$ groups $\Q^{(2)} \otimes \Q^{(2)}$ and $\Q^{(2)} \otimes \P_1$ are part of the local series
\[
    \Q^{(r_1)} \otimes \Q^{(r_2)}, \numberthis
\]
which has order $4 \cdot 2^{r_1 + r_2} \tau$ and is in the $\max(r_1,r_2)$ level of the Clifford hierarchy.

\subsection{Non-Monomial Imprimitive Groups in Higher Levels of the Clifford Hierarchy}

\subsubsection{Local}

The non-monomial groups $\P_1 \otimes \C_1$ and $\Q^{(2)} \otimes \C_1$ are part of the same series
\[
    \Q^{(r)} \otimes \C_1, \numberthis
\]
of order $48\cdot 2^r \tau$. These groups are in the $r$-th level of the Clifford hierarchy for $r \geq 2$ (and the $r=1$ case is in the Clifford group). 

The non-monomial groups $\P_1 \otimes \C_1'$ and $\Q^{(2)} \otimes \C_1'$ are part of the same series
\[
    \Q^{(r)} \otimes \C_1', \numberthis
\]
of order $24\cdot 2^r \tau$. These groups are in the $r$-th level of the Clifford hierarchy for $r \geq 2$ (and the $r=1$ case is in the Clifford group).

\subsubsection{Entangling}

The group $\N(768\sigma) = \expval{\Q^{(2)} \otimes \C_1, \CNOT}$ is part of the series
\[
    \N(192\cdot 2^r\sigma) := \expval{\Q^{(r)} \otimes \C_1, \CNOT}. \numberthis
\]
These groups are in the $r$-th level of the Clifford hierarchy for $r \geq 2$ (and the $r=1$ case is in the Clifford group). 

Lastly recall the group $\N(384\sigma)_a = \expval{\Q^{(2)} \otimes \C_1',\CNOT}$. This is  part of the series
\[
    \N(96\cdot 2^r\sigma) := \expval{\Q^{(r)} \otimes \C_1', \CNOT}. \numberthis
\]
These groups are in the $r$-th level of the Clifford hierarchy for $r \geq 2$ (and the $r=1$ case is in the Clifford group).

\newpage 
\onecolumngrid
\section{Table Of All Primitive Finite Subgroups of $\SU(4)$}

For convenience, here we list all 31 primitive subgroups of $\SU(4)$. Although 14 of these groups are irrelevant to stabilizer codes (since they are either exotic or do not contain the Pauli group), we list all of these groups here as an update to \cite{subgroupsSU4} which contains several errors. Recall that the projective group is just $\G$ mod scalar matrices like $ i\I\I$.

\begin{table*}[htp]
    \centering
    \begin{tabular}{|c||c|c|c|c|c|c|c|} \hline   \cite{subgroupsSU4}  & Class Name & Group Name &  Quantum Gates & Order & Projective Group \\ \hline \hline 
        $10$ & \multirow{4}{*}{\shortstack{Clifford \\ local}} & $ \C_1' \otimes \C_1'$ & $\expval{\Z  \I, \F \I, \I \Z, \I \F }$ & $288=144 \tau $ & \texttt{SmallGroup(144,184)}\\
        $11$ & & $ \C_2^{\bowtie'} $ & $\expval{\H  \S, \S  \H, \F \F }$ & $576=288 \tau $ & \texttt{SmallGroup(288,1026)} \\
        $12$ & & $ \C_1' \otimes \C_1$ & $\expval{\Z  \I, \F \I, \I \S, \I \H }$ & $576=288\tau$ & \texttt{SmallGroup(288,1024)}\\ 
        $14$ & & $ \C_1 \otimes \C_1 $ & $\expval{\S  \I, \H \I, \I \S, \I \H }$ & $1152=576 \tau $ & \texttt{SmallGroup(576,8653)}  \\ \hline 
        $17$ & \multirow{4}{*}{\shortstack{Clifford \\ non-entangling}} & $ \C(576\sigma)_{\ZZ[\zeta_8]}$ & $\expval{\H  \S, \F \F , \mathsf{SWAP} }$ & $2304=576\sigma$ & \texttt{SmallGroup(576,8654)} \\ 
        $18$ & & $ \C(576\sigma)_{\ZZ[i]}$ & $\expval{ \H  \S, \F \F , \S \I \cdot \Swap }$& $2304=576 \sigma $  & \texttt{SmallGroup(576,8652)} \\ 
        $19$ & & $ \C_1' \bowtie \C_1' $ & $\expval{\Z \I,\F \I, \mathsf{SWAP}}$ &  $1152=288\sigma$ & \texttt{SmallGroup(288,1025)} \\ 
        $21$ & & $ \C_1 \bowtie \C_1 $ & $\expval{\S \I , \H \I, \mathsf{SWAP}}$  & $4608=1152 \sigma $ & \texttt{SmallGroup(1152,157849)} \\ \hline 
        $22$ & \multirow{9}{*}{\shortstack{Clifford \\ entangling \\ with $ \P_2 $}} & $ \C(80\sigma) $ & $\expval{ \P_2, \mathsf{BELL} }$ & $320=80 \sigma $ & \texttt{SmallGroup(80,49)}  \\ 
        $23$ & & $ \C(160\sigma) $ & $\expval{ \P_2, \K^2, \mathsf{BELL} }$  & $640=160 \sigma $ & \texttt{SmallGroup(160,234)} \\ 
        $24$ & & $ \C(320\sigma) $ & $\expval{ \P_2, \K, \mathsf{BELL} }$ & $1280=320 \sigma $ & \texttt{SmallGroup(320,1635)} \\ 
        $25$ & &  $ \C(960\sigma) $ & $ \expval{ \C_1' \otimes \P_1, \mathsf{BELL}} $ & $ 3840=960 \sigma $ & \texttt{PerfectGroup(960,2)} \\ 
        $26$ & &  $ \C_1(\mathbb{F}_4) $ & $\expval{ \P_2 , \mathsf{H} \mathsf{H}, \mathsf{BELL}  }$ & $3840=960\sigma $ & \texttt{PerfectGroup(960,1)}  \\
        $27$ &  & $ \C(1920 \sigma)_{\ZZ[i,\sqrt{2}]}$ &  $\expval{ \C_1 \otimes \P_1, \mathsf{BELL} }$ & $ 7680= 1920 \sigma $ & \texttt{SmallGroup(1920,240996)}  \\
        $28$ & & $ \C(1920 \sigma)_{\ZZ[\zeta_8]}$  & $\expval{ \P_2 , \mathsf{SWAP} , \mathsf{BELL}  }$   & $7680=1920\sigma $ & \texttt{SmallGroup(1920,240993)} \\ 
        $29$ &  & $ \C_2'$ & $\expval{ \C_1' \otimes \C_1', \mathsf{BELL} }$ & $23040=5760\sigma$ & \texttt{PerfectGroup(5760)} \\ 
        $30$ &  &  $\mathsf{C}_2 $ & $\expval{ \C_1 \otimes \C_1, \mathsf{BELL} } $ & $46080=11520\sigma $ & $-$ \\ \hline 
        $1$ & \multirow{6}{*}{} &  $\C(120\tau)_{\ZZ[i]}'$ &  $\expval{\mathsf{A}, \CZ \cdot \S\F}$ & $120=60 \tau $  & \texttt{PerfectGroup(60)} \\ 
      $2$ & \multirow{1}{*}{\shortstack{Clifford \\entangling \\ without $ \P_2 $}} & $\C(120\sigma)_{\ZZ[\zeta_8]}' $  & $\expval{\CZ^\dagger \cdot \Swap ,\BELL}$  & $60 $ & \texttt{PerfectGroup(60)} \\ 
      $3$ &  & $\C(360 \tau) $  & $\expval{\I \F, \BELL}$ & $ 720=360 \tau$  & \texttt{PerfectGroup(360)}  \\ 
      $7$ &  & $ \C(120\tau)_{\ZZ[i]}$ &   $\expval{\H\Z \cdot \mathsf{BELL}, \S\Z \cdot \mathsf{BELL} }$ & $ 240=120 \tau $  & \texttt{SymmetricGroup(5)} \\ 
      $8$ & & $\C(120\sigma)_{\ZZ[\zeta_8]}$ &  $\expval{\mathsf{SWAP},\mathsf{BELL} }$  & $ 480=120\sigma $ & \texttt{SymmetricGroup(5)} \\ \hline 
      $13$ & \multirow{3}{*}{\shortstack{Exotic \\ local}} & $ \ico \otimes \C_1'$ & $\expval{ \Z  \I, \mathsf{\Phi}  \I, \I \Z,\I  \F}$ & $1440=720 \tau $ & \texttt{SmallGroup(720,768)} \\
        $15$ & & $ \ico \otimes \C_1$ & $\expval{ \Z  \I, \mathsf{\Phi}  \I, \I \S,\I  \H}$ & $2880=1440\tau$ & \texttt{SmallGroup(1440,5848)}  \\ 
        $16$ & & $ \ico \otimes \ico $ & $\expval{ \Z  \I, \mathsf{\Phi}  \I, \I \Z,\I  \mathsf{\Phi}}$ & $7200=3600 \tau $ & \texttt{PerfectGroup(3600)}  \\ \hline
        $20$ & {\shortstack{Exotic $ \bowtie $}} & $ \ico \bowtie \ico $ & $\expval{ \Z  \I, \mathsf{\Phi}  \I, \Swap}$ & $28800=7200 \sigma $ & $-$  \\ \hline
                $4$ & \multirow{6}{*}{} & $\Ex(2520\tau)$ & $\expval{\mathsf{V}_1,\mathsf{V}_2,\DCNOT}$ & $ 5040=2520 \tau $  & \texttt{PerfectGroup(2520)} \\ 
      $5$ & \multirow{3}{*}{\shortstack{Exotic \\entangling}} & $\Ex(168\tau) $  & $\expval{\mathsf{V}_1,\mathsf{V}_2}$ & $ 336=168 \tau $ & \texttt{PerfectGroup(168)} \\ 
      $6$ &  & $\Ex(25920 \tau) $  & $ \expval{\mathsf{U}_1,\mathsf{U}_2,\DCNOT}$ & $ 51840=25920 \tau$  & \texttt{PerfectGroup(25920)}\\
      $9$ & & $\Ex(720\sigma)$ & $\expval{\mathsf{W}_1,\mathsf{W}_2,\mathsf{W}_3, \Z\X, \Z\X }$  & $ 2880=720\sigma $ & \texttt{SymmetricGroup(6)} \\
      $7$ &  & $ \Ex(120\sigma)$ &  $\expval{\mathsf{W}_1,\mathsf{W}_2,\mathsf{W}_3, \Z\X \cdot \Z\X}$  & $ 480=120 \sigma $ & \texttt{SymmetricGroup(5)}  \\ 
       \hline
    \end{tabular}
    \caption{The 31 primitive subgroups of $\SU(4)$. There are only 30 subgroups up to isomorphism because $ \C(120\sigma)_{\ZZ[i]} \cong \Ex(120\sigma) $, but these subgroups are strongly inequivalent since their characters generate different rings.}
\end{table*}

\ \newpage 
\section{The 56 subgroups of $\C_2$ that contain $\P_2$ }

\begin{table*}[htp]
    \centering
    \scalebox{0.9}{
    \begin{tabular}{c|c|c|c|c|c|c} \toprule    \cite{subgroupsSU4}  & Class Name & Group Name &  Quantum Gates & Order & GAP & Projective Group GAP \\ \toprule  
        $10$ & \multirow{4}{*}{\shortstack{Primitive \\ local}} & $ \C_1' \otimes \C_1'$ & $\expval*{\Z  \I, \F \I, \I \Z, \I \F }$ & $288=144 \tau $ & [288,860] & [144,184]\\
        
        $11$ & & $ \C_2^{\bowtie'} $ & $\expval*{\H  \S, \S  \H, \F \F }$ & $576=288 \tau $ & [576, 8282] & [288,1026] \\
        
        $12$ & & $ \C_1' \otimes \C_1$ & $\expval*{\Z  \I, \F \I, \I \S, \I \H }$ & $576=288\tau$ & [576, 8273] & [288,1024]\\
        
        $14$ & & $ \C_1 \otimes \C_1 $ & $\expval*{\S  \I, \H \I, \I \S, \I \H }$ & $1152=576 \tau $ & [1152, 157463] & [576,8653]  \\ \hline

        $17$ & \multirow{4}{*}{\shortstack{Primitive \\ Non-entangling}} & $ \C(576\sigma)_{\ZZ[\zeta_8]}$ & $\expval{\H  \S, \F \F , \mathsf{SWAP} }$ & $2304=576\sigma$ & $-$ & [576,8654] \\ 
        
        $18$ & & $ \C(576\sigma)_{\ZZ[i]}$ & $\expval{ \H  \S, \F \F , \S \I \cdot \Swap }$& $2304=576 \sigma $  & $-$  & [576,8652] \\
        
        $19$ & & $ \C_1' \bowtie \C_1' $ & $\expval{\Z \I,\F \I, \mathsf{SWAP}}$ &  $1152=288\sigma$ & [1152,155473] & [288,1025] \\
        
        $21$ & & $ \C_1 \bowtie \C_1 $ & $\expval{\S \I , \H \I, \mathsf{SWAP}}$  & $4608=1152 \sigma $ & $-$ & [1152,157849] \\ \hline

        $22$ & \multirow{9}{*}{\shortstack{Primitive \\ Entangling }} & $ \C(80\sigma) $ & $\expval{ \P_2, \mathsf{BELL} }$ & $320=80 \sigma $ & [320,1586] & [80,49]  \\
        
        $23$ & & $ \C(160\sigma) $ & $\expval{ \P_2, \K^2, \mathsf{BELL} }$  & $640=160 \sigma $ & [640, 21464] & [160,234] \\ 
        
        $24$ & & $ \C(320\sigma) $ & $\expval{ \P_2, \K, \mathsf{BELL} }$ & $1280=320 \sigma $ & [1280, 1116380] & [320,1635] \\
        
        $25$ & &  $ \C(960\sigma) $ & $ \expval{ \C_1' \otimes \P_1, \mathsf{BELL}} $ & $ 3840=960 \sigma $ & $-$ & (960,2) \\ 
        
        $26$ & &  $ \C_1(\mathbb{F}_4) $ & $\expval{ \P_2 , \mathsf{H} \mathsf{H}, \mathsf{BELL}  }$ & $3840=960\sigma $ & (3840,2) & (960,1)  \\
        
        $27$ &  & $ \C(1920 \sigma)_{\ZZ[i,\sqrt{2}]}$ &  $\expval{ \C_1 \otimes \P_1, \mathsf{BELL} }$ & $ 7680= 1920 \sigma $ & $-$ & [1920,240996]  \\
        
        $28$ & & $ \C(1920 \sigma)_{\ZZ[\zeta_8]}$  & $\expval{ \P_2 , \mathsf{SWAP} , \mathsf{BELL}  }$   & $7680=1920\sigma $ & $-$ & [1920,240993] \\ 
        
        $29$ &  & $ \C_2'$ & $\expval{ \C_1' \otimes \C_1', \mathsf{BELL} }$ & $23040=5760\sigma$ & (23040,2) & (5760,1) \\ 
        
        $30$ &  &  $\mathsf{C}_2 $ & $\expval{ \C_1 \otimes \C_1, \mathsf{BELL} } $ & $46080=11520\sigma $ & $-$ & $-$ \\ \hline

         & \multirow{6}{*}{\shortstack{Imprimitive \\ Monomial \\ Shape $ S_4 $}} & $\M(768\sigma,S_4)$ & $\expval*{\Q^{(2)} \otimes \Q^{(2)}, \CNOT_{12}, \CNOT_{21}  }$ & $3072=768\sigma$ & $-$ & [768, 1090135]
 \\
         
         & & $\M(384\sigma,S_4)_{\mathbb{Z}[i]}$ & $\expval*{\P_2, \CNOT_{12} \cdot \S\I, \CNOT_{21} \cdot \S\I }$ & $1536=384\sigma$ & [1536,408569063] & [384, 18135] \\

        & & $\M(384\sigma,S_4)_{\mathbb{Z}[i,\sqrt{2}]}$ & $\expval*{\P_2 , \CNOT_{12}, \CNOT_{21} , \S\S  }$ & $1536=384\sigma$ & [1536,408569058] & [384, 18135]
 \\

        & & $\M(192\sigma,S_4)$ & $\expval*{\P_2,\CNOT_{12}, \CNOT_{21}, \CZ}$ & $768 = 192\sigma$ & [768, 1085977] & [192,955] \\
    
        & & $\M(96\sigma,S_4)_{\mathbb{Z}[i]}$ & $\expval*{\P_2 ,\CNOT_{12} \cdot \S\I, \CNOT_{21} \cdot \I\S }$ & $384 = 96\sigma$ & [384,20096] & [96,227] \\
    
         & & $\M(96\sigma,S_4)_{\mathbb{Z}[i,2\zeta_8]}$ & $\expval*{ \P_2, \CNOT_{12}, \CNOT_{21} }$ & $384 = 96\sigma$ & [384,18142]
& [96,227] \\  \hline

         & \multirow{4}{*}{\shortstack{Imprimitive \\ Monomial \\ Shape $ A_4 $}} & $\M(384\sigma,A_4)$  & $\expval*{\Q^{(2)} \otimes \Q^{(2)}, \DCNOT }$ & $1536 = 384\sigma$ & [1536,408535094] & [384, 18236]
  \\
        
         && $\M(192\sigma,A_4)$ & $\expval*{\P_2, \DCNOT, \S\S  }$ & $768 = 192\sigma$ & [768,1083945] & [192,1023] \\
         
         && $\M(96\sigma,A_4)$ & $\expval*{\P_2 ,\DCNOT, \CZ}$ & $384 = 96\sigma$ & [384,603] & [96,70] \\
         
         && $\M(48\sigma,A_4)$ & $\expval*{\P_2, \DCNOT}$ & $192 = 48\sigma$ & [192,1509] & [48,50] \\
         \hline

         & \multirow{13}{*}{\shortstack{Imprimitive \\ Monomial \\ Shape $ D_4 $}} & $\M(256\sigma,D_4)$ & $\expval*{\Q^{(2)} \otimes \Q^{(2)}, \CNOT }$ & $1024=256\sigma$ & $-$  & [256, 26547] \\
    
        && $\M(128\sigma,D_4)_a$ & $\expval*{\Q^{(2)} \otimes \P_1, \CNOT, \CZ }$  & $512=128\sigma$ & [512,419131] & [128,1759]  \\
    
        && $\M(128\sigma,D_4)_b$ & $\expval*{\Q^{(2)} \otimes \P_1, \CNOT \cdot \I\S }$ & $512=128\sigma$ & [512,60109] & [128,854] \\
    
        && $\M(128\sigma,D_4)_c$ & $\expval*{\P_1 \otimes \Q^{(2)},\CNOT }$ & $512=128\sigma$ & [512,59383] & [128,928] \\
    
        && $\M(128\sigma,D_4)_d$ & $ \expval*{\P_1 \otimes \Q^{(2)}, \CNOT \cdot \S\I, \CZ \cdot \S\I  }$  & $512=128\sigma$ & [512,420089] & [128,928] \\
    
        && $\M(128\sigma,D_4)_e$ & $\expval*{\P_2,\CNOT, \S\S }$ & $512=128\sigma$ & [512,60476] & [128,931]  \\
    
        && $\M(128\sigma,D_4)_f$ & $\expval*{\P_2, \CNOT \cdot \S\I , \S\S }$  & $512=128\sigma$ & [512,60321] & [128,931]  \\ 

        && $\M(64\sigma,D_4)_a$ & $\expval*{ \Q^{(2)} \otimes \P_1, \CNOT }$  & $256=64\sigma$ & [256,17275] & [64,216] \\
    
        && $\M(64\sigma,D_4)_b$ & $\expval*{\P_2 ,\CNOT, \CZ}$  & $256=64\sigma$ & [256,6039] & [64,138] \\
    
        && $\M(64\sigma,D_4)_c$ & $\expval*{\P_2, \CNOT \cdot \I \S,  \CZ \cdot \S \I }$  & $256=64\sigma$ & [256,6552] & [64,32] \\
     
        && $\M(64\sigma,D_4)_d$ & $\expval*{\P_2, \CNOT \cdot \S \S,  \CZ \cdot \S\I }$ & $256=64\sigma$ & [256,6560] & [64,32]  \\
    
        && $\M(64\sigma,D_4)_e$ & $ \expval*{\P_2, \CNOT \cdot \S\I, \CZ \cdot \S\I  }$ & $256=64\sigma$ & [256,26555] & [64,138] \\

        && $\M(32\sigma,D_4)$ & $\expval*{\P_2, \CNOT }$ & $128=32\sigma$ & [128,523] & [32,27] \\ \hline

        & \multirow{6}{*}{\shortstack{Imprimitive \\ Monomial \\ Shape $ V_4 $}} & $\M(128\sigma,V_4)$ & $\expval*{\Q^{(2)} \otimes \Q^{(2)}, \CZ }$ & $512=128\sigma$ & [512,7521281] & [128,2281] \\

        && $\M(64\sigma,V_4)$ & $ \expval*{\P_2, \S\S, \CZ\cdot \S\I  }$ & $256=64\sigma$ & [256,24064] & [64,242] \\

        && $\M(32\sigma,V_4)$ & $ \expval*{\P_2, \CZ\cdot \S\I } $ & $128=32\sigma$ & [128,1750] & [32,27] \\

        && $\Q^{(2)} \otimes \Q^{(2)}$ & $\expval*{\X\I,\S\I,\I\X,\I\S }$ & $128=64\tau$ & [128,2024] & [64,226]\\

        && $\Q^{(2)} \otimes \P_1$ & $ \expval*{\X\I, \S\I, \I\X,\I\Z } $ & $64=32\tau$ & [64,257] & [32,46] \\

        && $\P_2$ & $ \expval*{ \X\I,\Z\I, \I\X,\I\Z }$ & $32=16\tau$ & [32,49]  & [16,14] \\ \hline

        & \multirow{5}{*}{\shortstack{Imprimitive \\ Non-Monomial \\ Local}} & $\Q^{(2)} \otimes \C_1$ & $\expval{\X\I,\S\I,\I\S,\I\H}$  & $384=192\tau$ & [384,18044] & [192,1472] \\
        
        && $\P_1 \otimes \C_1$ & $\expval{\X\I,\Z\I,\I\S,\I\H}$ & $192=96\tau$ & [192,1484] & [96,226] \\
        
        && $\Q^{(2)} \otimes \C_1'$ & $\expval{\X\I,\S\I,\I\X,\I\F}$ & $192=96\tau$ & [192,1017] & [96,197] \\
        
        && $\P_1 \otimes \C_1'$ & $\expval{\X\I,\Z\I,\I\X,\I\F}$ & $96=48\tau$ & [96,201] & [48,49] \\

        && $\N(96\tau)$ & $ \expval{ \P_1 \otimes \C_1', \S \S  }$ & $192 = 96\tau$ & [192,988] & [96,195] \\
        
         \hline 

         & \multirow{6}{*}{\shortstack{Imprimitive \\ Non-Monomial \\ Entangling}} &
         $\N(768\sigma)$ & $\expval*{\Q^{(2)} \otimes \C_1, \CNOT}$ & $3072 = 768\sigma$ & $-$ & [768, 1088565] \\
         
        && $\N(384\sigma)_a$ & $\expval*{\Q^{(2)} \otimes \C_1', \CNOT }$ & $1536 = 384\sigma$ & [1536,408528836] & [384,5837] \\
        
        && $\N(384\sigma)_b$ & $\expval*{\P_1 \otimes \C_1, \CZ\cdot \S\I }$ & $1536=384\sigma$ & [1536,408557064] & [384,5602]  \\
        
        && $\N(384\sigma)_c$  & $\expval*{ \P_1 \otimes \C_1', \S \S , \CZ \cdot \S\I }$ & $1536 = 384\sigma$ & [1536,408546526] & [384,5603] \\
        
        && $\N(192\sigma)$ & $\expval*{\P_1 \otimes \C_1', \CZ\cdot \S\I }$ & $768=192\sigma$ & [768,1084090] & [192,201] \\
        
         \bottomrule
    \end{tabular}
    }
    \caption{Square brackets are \texttt{SmallGroup} IDs and parenthesis are \texttt{PerfectGroup} IDs in GAP}
\end{table*}

\end{document}